%% file: main.tex
\documentclass[manuscript,screen]{acmart}
\AtBeginDocument{%
  }

\usepackage{mathrsfs}
\usepackage{subfigure}
\usepackage{microtype,flushend}
\usepackage{colortbl}
\usepackage{balance}
\usepackage[most]{tcolorbox}
\usepackage{multirow}
\usepackage{fontawesome}
\usepackage{threeparttable}
\usepackage{color}
\usepackage[ruled,vlined]{algorithm2e}
\usepackage{xspace}
 \setcopyright{none}
\renewcommand\footnotetextcopyrightpermission[1]{}

\newcommand{\tabincell}[2]{\begin{tabular}{@{}#1@{}}#2\end{tabular}}

\usepackage[framemethod=tikz]{mdframed}
\newcounter{finding}
\newcommand{\finding}[1]{\refstepcounter{finding}
  \vspace{2.3mm}
 \begin{mdframed}[linecolor=gray,roundcorner=12pt,backgroundcolor=gray!15,linewidth=3pt,innerleftmargin=2pt, leftmargin=0cm,rightmargin=0cm,topline=false,bottomline=false,rightline = false]
  \textbf{Ans. to RQ\arabic{finding}:} #1
 \end{mdframed}
 \vspace{2.3mm}
}

\lstdefinestyle{mystyle}{
    numberstyle=\tiny,
    basicstyle=\ttfamily\footnotesize,
    breakatwhitespace=false,         
    breaklines=true,                 
    captionpos=b,                    
    keepspaces=true,                 
    numbers=left,                    
    numbersep=5pt,                  
    showspaces=false,                
    showstringspaces=false,
    showtabs=false,                  
    tabsize=2,
    frame={bottomline}
}
\setcopyright{acmcopyright}
\copyrightyear{2018}
\acmYear{2018}
\acmDOI{XXXXXXX.XXXXXXX}

\begin{document}

\title{\toolName: General Application-Level Cold-Start Latency Optimization for Function-as-a-Service in Serverless Computing}


\author{Xuanzhe Liu}
\affiliation{%
  \institution{Peking University}
  \city{Beijing}
  \country{China}}
\email{liuxuanzhe@pku.edu.cn}

\author{Jinfeng Wen}
\affiliation{%
  \institution{Peking University}
  \city{Beijing}
  \country{China}}
\email{jinfeng.wen@stu.pku.edu.cn}

\author{Zhenpeng Chen}
\affiliation{%
  \institution{University College London}
  \city{London}
  \country{United Kingdom}}
\email{zp.chen@ucl.ac.uk}

\author{Ding Li}
\affiliation{%
  \institution{Peking University}
  \city{Beijing}
  \country{China}}
\email{ding_li@pku.edu.cn}

\author{Junkai Chen}
\affiliation{%
  \institution{Peking University}
  \city{Beijing}
  \country{China}}
\email{cjk@pku.edu.cn}

\author{Yi Liu}
\affiliation{%
  \institution{Advanced Institute of Big Data, Beijing}
  \city{Beijing}
  \country{China}}
\email{liuyi14@pku.edu.cn}

\author{Haoyu Wang}
\affiliation{%
  \institution{Huazhong University of Science and Technology}
  \city{Hubei}
  \country{China}}
\email{haoyuwang@hust.edu.cn}

\author{Xin Jin}
\affiliation{%
  \institution{Peking University}
  \city{Beijing}
  \country{China}}
\email{xinjinpku@pku.edu.cn}







\renewcommand{\shortauthors}{Liu et al.}
\newcommand{\para}[1]{\smallskip\noindent{\bf {#1}. }}
\newcommand{\toolName}{\textit{FaaSLight}\xspace}
\begin{abstract}
\input{section/abstract}
\end{abstract}



\maketitle

\input{section/introduction}

\input{section/motivationnew}

\input{section/design}
\input{section/evaluation}
\input{section/result}

\input{section/discussion}
\input{section/relatedwork}

\input{section/conclusion}

\begin{acks}
This work was supported by the National Key Research and Development Program of China (No. 2020YFB2104100), the National Natural Science Foundation of China under the grant numbers 62172008 and 62172009, and the National Natural Science Fund for the Excellent Young Scientists Fund Program (Overseas). Zhenpeng Chen is supported by the ERC Advanced Grant under the grant number 741278 (EPIC: Evolutionary Program Improvement Collaborators).

We thank the valuable review comments from anonymous reviewers and the timely and kindly review process organized by the associate editor, Prof. Sam Malek. If any concerns are raised, please contact zp.chen@ucl.ac.uk and ding\_li@pku.edu.cn for correspondence.

\end{acks}



\bibliographystyle{ACM-Reference-Format}
\bibliography{main}


\end{document}

%% file: section/abstract.tex
Serverless computing is a popular cloud computing paradigm that frees developers from server management. Function-as-a-Service (FaaS) is the most popular implementation of serverless computing, representing applications as event-driven and stateless functions. However, existing studies report that functions of FaaS applications severely suffer from cold-start latency. In this paper, we propose an approach namely \toolName to accelerating the cold start for FaaS applications through application-level optimization. 
We first conduct a measurement study to investigate the possible root cause of the cold start problem of FaaS. The result shows that application code loading latency is a significant overhead. Therefore, loading only indispensable code from FaaS applications can be an adequate solution. Based on this insight, we identify code related to application functionalities by constructing the function-level call graph, and separate other code (i.e., optional code) from FaaS applications. The separated optional code can be loaded on demand to avoid the inaccurate identification of indispensable code causing application failure. In particular, a key principle guiding the design of \toolName is inherently general, i.e., \textit{\textbf{platform}}- and \textit{\textbf{language-agnostic}}. In practice, \toolName can be effectively applied to FaaS applications developed in different programming languages (Python and JavaScript), and can be  seamlessly deployed on popular serverless platforms such as AWS Lambda and Google Cloud Functions, without having to modify the underlying OSes or hypervisors, nor introducing any additional manual engineering efforts to developers. The evaluation results on real-world FaaS applications show that \toolName can significantly reduce the code loading latency (up to 78.95\%, 28.78\% on average), thereby reducing the cold-start latency. As a result, the total response latency of functions can be decreased by up to 42.05\% (19.21\% on average). Compared with the state-of-the-art, \toolName achieves a \textbf{21.25$\times$} improvement in reducing the average total response latency.

%% file: section/introduction.tex
\section{Introduction}\label{intro}


Serverless computing is a popular cloud computing paradigm and has been applied to various domains, including machine learning~\cite{feng2018exploring}, scientific computing~\cite{shankar2020numpywren}, and video processing~\cite{ao2018sprocket}. It is predicted that 50\% of global enterprises will employ serverless computing by 2025~\cite{gartner20new}. To embrace this paradigm, major cloud vendors have rolled out various serverless platforms, such as AWS Lambda~\cite{aws}, Microsoft Azure Functions~\cite{azure}, and Google Cloud Functions~\cite{google}. In these serverless platforms, Function-as-a-Service (FaaS) is the most prominent implementation pattern~\cite{AkkusATC18, mohanty2018evaluation, taibi2020serverless, JonasCoRR2019new, tariq2020sequoia, wen2023rise}. FaaS represents event-driven and stateless functions (\textit{serverless functions}). Developers can implement their applications as a combination of serverless functions, and these applications are called FaaS applications.
The underlying serverless platforms automatically handle resource management. Therefore, developers do not need to manage servers or VM instances to run FaaS applications. Instead, serverless functions are dynamically allocated with resources when they are triggered by events (e.g., an HTTP request). If a serverless function has not been used for a while, the platform will release the resources. In this way, resource management can be lightweight and efficient.



However, such on-demand resource management in serverless computing introduces the cost of longer response latency. The resources of idle serverless functions will be released. Therefore, the serverless platform has to initialize the execution environment for the invocations to functions that are not frequently used, which prolongs the functions' response latency. In practice, the latency spent on preparing the execution environment (cold-start latency for short) has been demonstrated to be significant in the total response latency in various application scenarios, such as IoT applications~\cite{pelle2020operating, zhang2021edge}, video processing~\cite{ao2018sprocket, fouladi2017encoding}, and machine learning~\cite{carreira2019case, feng2018exploring}. For example, a previous 
study~\cite{fuerst2021faascache} reported that the cold-start latency could be as much as 80\% of the total response latency. Therefore, optimizing cold-start latency is a critical challenge of contemporary FaaS applications.


Several efforts have been made to optimize the cold-start latency at the \textit{system level} (i.e., optimizing the underlying serverless platforms), such as developing lightweight virtualization technology of containers~\cite{DuASPLOS2020}, adjusting the scheduling policy to keep instances warm~\cite{WangHW19}, and redesigning sandbox runtime mechanisms~\cite{OakesATC18,AkkusATC18}. Although these efforts are demonstrated to be efficient and promising, they all inherently require extensive engineering efforts to modify underlying OSes or hypervisors. Serverless platform vendors should have concerns about adopting and implementing substantial changes to their existing infrastructures. In addition, they also have concerns about security mechanisms, e.g., ASLR~\cite{DuASPLOS2020}. To the best of our knowledge, none of the aforementioned techniques have been applied to commercial serverless platforms.

In this paper, we focus on applications built on FaaS. Compared to system-level optimization of cold-start latency, which requires intrusive changes to the underlying platforms, we aim to tackle this problem at the application level. Our guiding principle is to provide a \textit{platform/language-independent} and \textit{developer-free} technique that application developers can adopt to optimize the cold-start latency of serverless functions on most existing platforms. To achieve our goal, we first investigate the possible root cause of the cold start overhead of FaaS applications. We conduct a measurement study on real-world FaaS applications. The result shows that the application code loading latency takes a significant part of the cold-start latency.

Based on this observation, we propose an application-level approach namely \toolName to optimizing the cold-start latency of serverless functions by reducing the size of executed code, i.e., loading only necessary code.
One key insight of \toolName is to identify code related to application functionalities (called \emph{indispensable code}) through constructing the function-level call graph. Then, it separates other code (called \emph{optional code}) from the original application by analyzing the intermediate representation of the application code. To guarantee the correctness of the application, \toolName does not remove the optional code, but compresses the optional code into a lightweight file and fetches the separated code in an on-demand loading way if the code is invoked. In this way, we can reduce the code size in the loading process and guarantee the correctness and availability of the final FaaS applications.

It is worth mentioning two key design principles raised by \toolName. First, \toolName presents  the support of \textit{generalizability}. That is to say, the principled design behind \toolName itself is independent of underlying heterogeneous serverless platforms and programming languages of the FaaS. As demonstrated later in this paper, \toolName can be easily applied to FaaS applications that are developed in different programming languages (i.e., Python and JavaScript) and executed on different platforms (i.e., popular AWS Lambda and Google Cloud Functions). Second, \toolName is \textit{developer-free} to existing (or legacy) FaaS applications.
Inherently, \toolName does not introduce \textbf{any additional} manual efforts for developers, nor requires any modification of existing serverless platforms. In practice, \toolName can be simply deployed as a service on current serverless platforms. When developers upload their applications, \toolName can process the code optimization and application deployment automatically. In other words, developers are unaware of any changes when \toolName works. 

We implement \toolName as the Python and JavaScript prototypes since they are the most widely used languages in the serverless community. We evaluate its effectiveness on real-world FaaS applications. The results show that \toolName can reduce the application code loading latency by up to 78.95\% (on average 28.78\%), thereby reducing the cold-start latency. As a result, the total response latency of serverless functions can be decreased by up to 42.05\% (on average 19.21\%). As an additional benefit, \toolName can decrease the runtime memory of serverless functions by up to 58.82\% (on average 14.79\%) due to the reduced size of loaded code. Compared with the state-of-the-art, \toolName achieves a 21.25$\times$ improvement in reducing the average total response latency.






To the best of our knowledge, \toolName is the first application-level effort to optimize the cold-start latency of serverless functions. To summarize, this paper makes the following contributions.

\begin{itemize}

\item We conduct a measurement study to demystify the possible root cause of the cold-start latency of serverless functions. The result shows that the application code loading latency is the significant overhead.

\item We propose a general application-level performance optimization approach to reducing the cold-start latency without compromising the effectiveness, correctness, and availability of FaaS applications.

\item We evaluate our approach using real-world FaaS applications under heterogeneous settings (different programming languages and platforms), and the results show that it can significantly reduce cold-start latency. We demonstrate that our approach, i.e., loading only indispensable code, can significantly improve the performance of FaaS applications.



\end{itemize}

We have implemented the prototype of \toolName to support both Python and JavaScript. The code and scripts have been open-sourced on GitHub\footnote{ \url{https://github.com/WenJinfeng/FaaSLight}}.

%% file: section/motivationnew.tex
\section{Background}\label{sec:background}

In this section, we introduce the background of serverless computing and then describe the cold start problem.

\subsection{Serverless Computing}




Serverless computing allows software developers to efficiently develop and deploy applications to the market without having to manage the underlying infrastructure~\cite{WangATC2018,wen2021characterizing,WenServerless21}, i.e, ``server-less'' means no server management for developers. In serverless computing, FaaS is the prominent and widely adopted implementation~\cite{WangATC2018, AkkusATC18, taibi2020serverless, JonasCoRR2019new, WenServerless21}. Developers can rely on the FaaS fashion to focus solely on the business logic of applications, which are composed of a set of serverless functions. Such applications are called FaaS applications.

In FaaS applications, serverless functions and their dependency libraries are packaged into a single bundle, and then deployed to serverless platforms. If the application size exceeds the deployment restriction (e.g., 250 MB uncompressed size on AWS Lambda), developers can deploy applications using container images with larger sizes~\cite{containerImageSupport}. After successful deployment, serverless functions will be triggered with predefined events, e.g., an HTTP request, a file update of cloud storage, or a timer going off. Once serverless functions are triggered, the serverless platform automatically allocates and launches dedicated function instances (e.g., VMs or containers) with restricted resources (e.g., CPU and memory) for them to execute their functionalities. When there are no incoming requests, launched instances and resources are later automatically released.





\begin{figure}[!thb]
	\centering
    \includegraphics[width=0.6\textwidth]{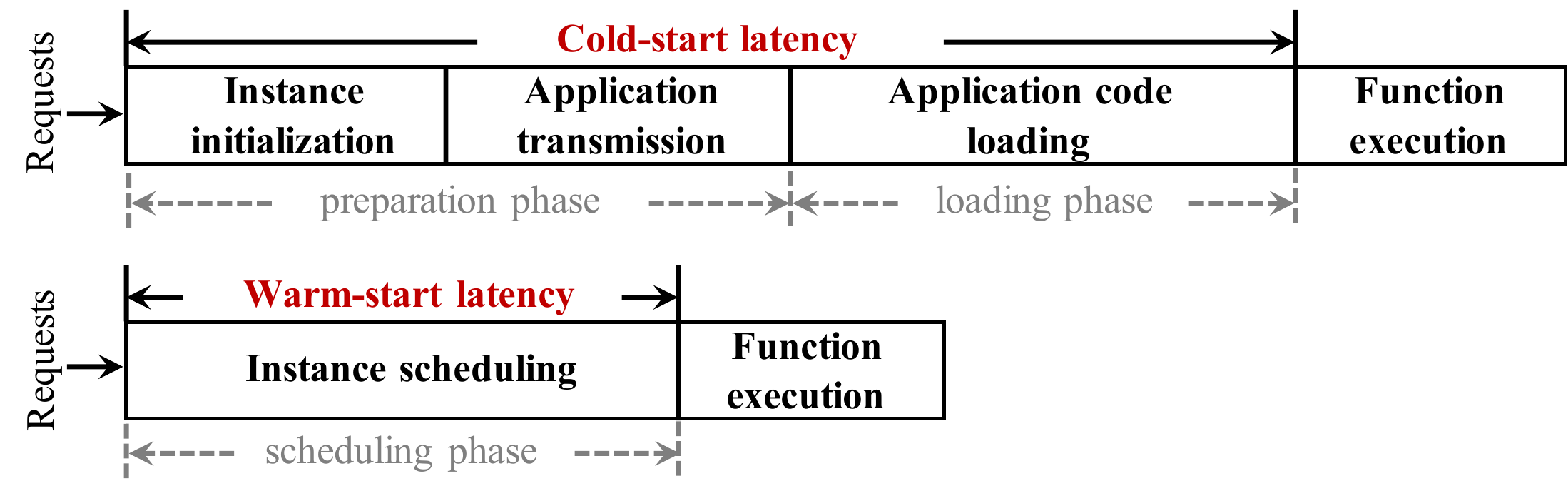}
    \caption{The latency breakdown after requests come in cold and warm starts.}
    \label{fig:latency}
\end{figure}

The invocation to a serverless function may go through two modes, the \textit{cold start mode} and the \textit{warm start mode}. If the invoked function has not been used for a threshold (keep-alive time), the invocation is in the cold start mode. In this mode, the serverless platform needs to prepare new VMs or containers, transmit the code of the function from remote cloud storage (such as AWS S3~\cite{yu2020characterizing, vahidinia2020cold}) to instances over the network, load the required code to initiate the application process, and finally execute the serverless function. On the contrary, if the invoked function is recently used (e.g., within 7 minutes for AWS Lambda~\cite{awsIdlelifetime}), the invocation is in the warm start mode, where the serverless platform reuses the launched instances of the same function. 

This paper focuses on the cold-start latency problem. For better illustration, we compare the cold-start latency with the warm-start latency in Fig.~\ref{fig:latency}.\footnote{Fig.~\ref{fig:latency} shows key latencies, not showing fine-grained latencies like request reception and return due to the black-box feature of commodity platforms.} The cold-start latency consists of three parts: the latency of preparing VMs or containers for the serverless function (instance initialization), transmitting the application over the network (application transmission), and loading the code related to the application (application code loading). We call instance initialization and application transmission as the preparation phase, and application code loading as the loading phase. 
In contrast, the warm-start latency includes only the latency of scheduling reused instances, which is called the scheduling phase in our study.

\subsection{The Cold Start Problem}



The cold-start latency significantly affects the overall runtime efficiency of a serverless function because (1) the cold start happens frequently~\cite{AkkusATC18,OakesATC18,ShahradATC20,fuerst2021faascache,wang2021faasnet}, and (2) once it happens, its latency takes a significant part of the end-to-end response latency of a serverless function~\cite{fuerst2021faascache,WangATC2018,DuASPLOS2020,yu2020characterizing,singhvi2021atoll,ShahradATC20}. Because serverless functions often execute in only a few milliseconds, cold-start latency is significant in comparison. Fuerst \textit{et al.}~\cite{fuerst2021faascache} showed that the cold-start latency could be as much as 80\% of the total response latency. Generally, serverless functions are short-lived. Du \textit{et al.}~\cite{DuASPLOS2020} calculated the ratio of function execution latency to total response latency for 14 serverless functions, and found that 12 serverless functions even cannot achieve 30\%, emphasizing that the total response latency of a serverless function is dominated by its startup time. Singhvi \textit{et al.}~\cite{singhvi2021atoll} also found that 57\% of serverless functions have an execution time of less than 100 ms. Therefore, a fast cold start is critical for developers because their tasks are often short-lived and completed quickly~\cite{HellersteinCIDR19,DuASPLOS2020,KlimovicWSTPK18,LiuPKP19}.

\section{Measurement study}\label{sec:measurement}

In this paper, we focus on applications built on FaaS. A FaaS application is composed of serverless functions. To further investigate the possible root cause of the cold-start latency, we conduct a measurement study on real-world FaaS applications executed on AWS Lambda, which is the most popular and widely used serverless platform~\cite{AWSLambdaTop1, AWSLambdaTop2}.

\begin{table}[ht]
    \footnotesize
      \caption{The details of our benchmarks.}
      \label{tab:datainfo}
        \begin{tabular}{l|l|r|r|r|l}
        \hline 
         \rowcolor{gray!50}\tabincell{c}{\textbf{App ID}}& \textbf{Name}&\tabincell{c}{\textbf{Size (MB)}}&\tabincell{c}{\textbf{FC (k)}}&\tabincell{c}{\textbf{LoC (k)}}&\textbf{Description}\cr
         
         \hline
        App1 & image-resize~\cite{ImageResizeApp} & 64.40 & 9.66 & 21.20 & resize image and save to Boto3\\

         \hline
        App2 & lambda-pillow~\cite{HelloworldApp} & 85.91 & 11.56 & 30.68 & import Boto3 and Pillow to test\\  
        
        \hline
        App3 & lambda-pandas~\cite{HellopandasApp} &83.88 & 41.76 & 151.40 & use Pandas to converse data\\
        
        \hline
        
        App4 & scikit-assign~\cite{scikitassignApp} &147.71 & 46.34 & 160.60  & use Sklearn model to predict price\\
        
         \hline
        App5 & lxml-requests~\cite{lxmlApp} &25.26 & 6.57 & 47.18 & use lxml to parse HTML pages\\ 
        
        \hline
        App6 & pandas-numpy~\cite{pandasnumpyApp} &113.36 & 56.83 & 192.05 & use Pandas and Numpy to generate data\\
        
        \hline
        App7 & skimage-lambda~\cite{skimageApp} &261.47& 72.63 & 267.94 & download image and use Skimage to process\\
        
        \hline
        App8 & opencv-pil~\cite{opencvpilApp} &152.96 &35.20 & 93.71 & use OpenCV and PIL to process image \\

        \hline
        App9 & wine-ml-lambda~\cite{wineMLApp} &248.25 & 81.91 & 317.88 & train Sklearn model and predict wine quality\\
        
        \hline
        App10 & lightgbm-sklearn~\cite{LightGBMApp} & 221.70 & 56.45 & 216.92 & use LightGBM model to predict \\
        
        \hline
        App11 & sentiment-analysis~\cite{SentimentApp} & 240.92 & 73.45 & 280.46 & use Sklearn model to predict sentiment statement\\
        
        \hline
        App12 & tensorflow-lambda~\cite{TensorflowApp} & 1217.23 & 61.46 & 260.38 & use TensorFlow regression model to predict data \\
        
        \hline
        App13 & numpy-lambda~\cite{NUMPYApp} & 70.22 & 37.69 & 99.12 & use Numpy to converse matrix data \\
         
        \hline
        App14 & lambda-opencv~\cite{OpenCVApp} &228.70 & 31.91 & 74.88 & use OpenCV to get properties\\
        
        \hline
        App15 & question-answering~\cite{questionanswerApp} & 2083.58  & 57.92 & 197.13  & use Bert model to answer questions\\
        
        \hline
         
        \textbf{Max} &  & 2083.58 & 81.91  & 317.88 & \\ 
       \textbf{Mean} &  &  349.70  & 45.42 & 160.77 & \\ 
        
        \hline
        \end{tabular}
       
    \end{table}

\subsection{Benchmarks}

We select real-world FaaS applications from GitHub as benchmarks according to the following criteria. In the FaaS application, serverless functions and required dependency libraries are generally bundled together to deploy and execute. A FaaS application is selected when (1) it is written in Python, which is one of the most widely used languages in the serverless community~\cite{serverlesscommunitysurvey}, (2) its application code contains more than 20k lines of code, indicating a median and large application~\cite{romano2018multi}, 
(3) it has detailed instructions to guide us to execute it successfully, and (4) it is not a development tool, such as AWS SAM CLI~\cite{AWSASM}, which is a command-line interface for developing serverless functions. Our final benchmarks consist of 15 real-world FaaS applications, ranging from data processing to machine learning. Specific details are shown in Table~\ref{tab:datainfo}. In our study, for each FaaS application, its application size, the number of functions contained in the application, and the number of lines of code are denoted as \emph{Size}, \emph{FC}, and \emph{LoC}, respectively. \emph{FC} is calculated by analyzing the intermediate representation of code and recognizing the number of function definitions, while \emph{LoC} is calculated by counting executable statements excluding single-line, multi-line, and document comments. As part of the serverless function, used dependency libraries are also involved in the calculations of \emph{Size}, \emph{FC}, and \emph{LoC}. On average, our benchmarks have 349.70 MB \emph{Size}, 45.42k \emph{FC}, and 160.77k \emph{LoC}.

\subsection{Measurement Results}\label{sec:measureresult}

We execute these FaaS applications in cold starts, and then obtain the preparation phase latency, loading phase latency, function execution latency (as described in Fig.~\ref{fig:latency}), as well as total response latency. The calculation of these latencies is as follows:

\emph{$\bullet$ Loading phase latency} is the latency of the application code loading in cold starts. It can be extracted from the ``\texttt{Init Duration}'' attribute provided by AWS Lambda execution logs. It is also practical to use time breakpoints between import code statements.

\emph{$\bullet$ Preparation phase latency} is the latency of the instance initialization and application transmission in cold starts. By setting time checkpoints at the request start and the beginning of the code body of serverless functions, we can obtain the overall cold-start latency including the preparation phase latency and loading phase latency. The preparation phase latency is extracted by removing the loading phase latency from the overall cold-start latency.

\emph{$\bullet$ Function execution latency} is the latency of executing serverless functions. It can be extracted from the ``\texttt{Duration}'' attribute of AWS Lambda execution logs. Similarly, this latency can also be obtained by using time breakpoints at the beginning and end of serverless function code execution.

\emph{$\bullet$ Total response latency} is the latency from request sending to request completion. By setting time checkpoints at the beginning and end of the request, the time interval is calculated as the total response latency.

\begin{figure}[!thb]
	\centering
    \includegraphics[width=0.6\textwidth]{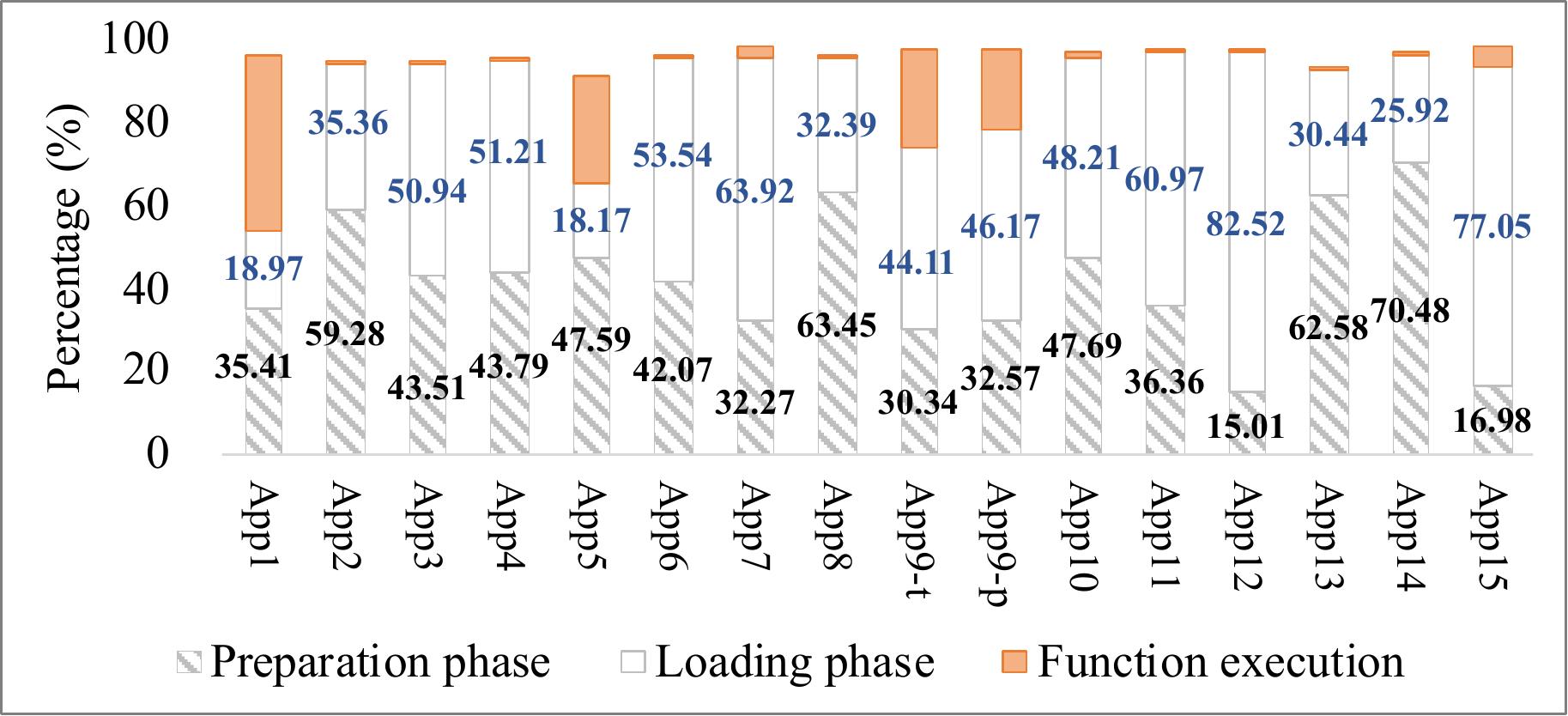}
    \caption{The percentage of each latency to the total response latency. 
    }
    \label{fig:cold_analysis}
\end{figure}

We report the percentage of each latency to the total response latency for tested FaaS applications in Fig.~\ref{fig:cold_analysis}.\footnote{App9 has two main functionalities, i.e., model training (App9-t) and model prediction (App9-p).} We observe that the loading phase latency is a significant overhead. Specifically, the cold-start latency, i.e., the sum of the preparation and loading phase latencies, takes up 88.70\% of the total response latency, while the function execution latency is only 7.57\%, on average. Particularly, for 12 FaaS applications, the function execution latency only takes less than 5\% of the total response latency. We further analyze the latency percentage of two phases in the cold-start latency. On average, the loading phase latency is 46.24\% of total response latency, while the preparation phase latency is 42.46\%, as can be seen from Fig.~\ref{fig:cold_analysis}. It illustrates that the application code loading latency is a significant overhead in cold starts. Generally, the preparation phase latency contains the latency used to prepare VMs or containers for the serverless function (instance initialization) and transmit the application over the network (application transmission). Developers can hardly control the preparation phase latency because commodity platforms such as AWS Lambda are not open to ordinary developers. The application code loading latency is caused by fetching application code in VMs or containers. In other words, reducing the code size of a FaaS application can optimize the application code loading latency.

FaaS applications are mostly written in high-level languages such as Python and JavaScript~\cite{serverlesscommunitysurvey}. In practice, third-party dependency libraries are often imported to help serverless functions implement complex functionalities. Although developers usually use only a small subset of all supported functionalities of these libraries, all functionalities are still loaded completely~\cite{quach2017multi,qian2019razor}. Thus, eliminating code not used by a FaaS application can potentially help optimize its application code loading latency. In practice, our previous experience demonstrated that a similar mechanism is effective to optimize client-side JavaScript  applications~\cite{Liu:SCIS14, Liu:SCW07}. The application code loading latency is a significant part of the cold-start latency, which further affects the end-to-end response latency. Therefore, optimizing the code size of a FaaS application can potentially reduce its overall end-to-end latency. This insight motivates our proposed approach in this paper.

%% file: section/design.tex
\section{Approach}\label{sec:design}


In this section, we propose the design of \toolName, which optimizes the cold-start latency of FaaS applications by eliminating the optional functions. We define optional functions as those that cannot be reached from the entries of a FaaS application. Thus, we can remove optional functions without causing runtime errors from the FaaS application. \toolName accepts the source code of an existing FaaS application and outputs a modified version of the input application with optional functions removed. The key challenge for \toolName is to guarantee the correctness of optional function elimination while maximizing the reduction of cold-start latency. Serverless functions contained in the FaaS application are often built with dynamic languages, such as Python and JavaScript~\cite{serverlesscommunitysurvey}. For such languages, it is particularly challenging, if not impossible, to identify the functions that can be reached from the entry of an invocation to the FaaS application with 100\% accuracy. An overly aggressive solution that aims to identify more optional functions may misclassify non-optional functions as optional, leading to crashes in the runtime. On the flip side, an overly conservative solution may miss optional functions in the elimination process, leading to sub-optimal cold start time reduction. To address this challenge, \toolName adopts a combined strategy. It first leverages a static-analysis-based technique that identifies the functions that are reachable from the entries of a FaaS application. We define the identified functions as indispensable functions. The static analysis of \toolName is aggressive that aims to identify as many indispensable functions as possible. Then, after the static analysis, \toolName adopts a conservative on-demand loading strategy, which does not directly delete the optional functions but replaces them with an on-demand loader. Thus, even if \toolName eliminates a non-optional function by mistake, the on-demand loader can still load the code of the eliminated function when it is invoked, ensuring the correctness of the FaaS application.



\begin{figure*}[!thb]
	\centering
    \includegraphics[width=0.9\textwidth]{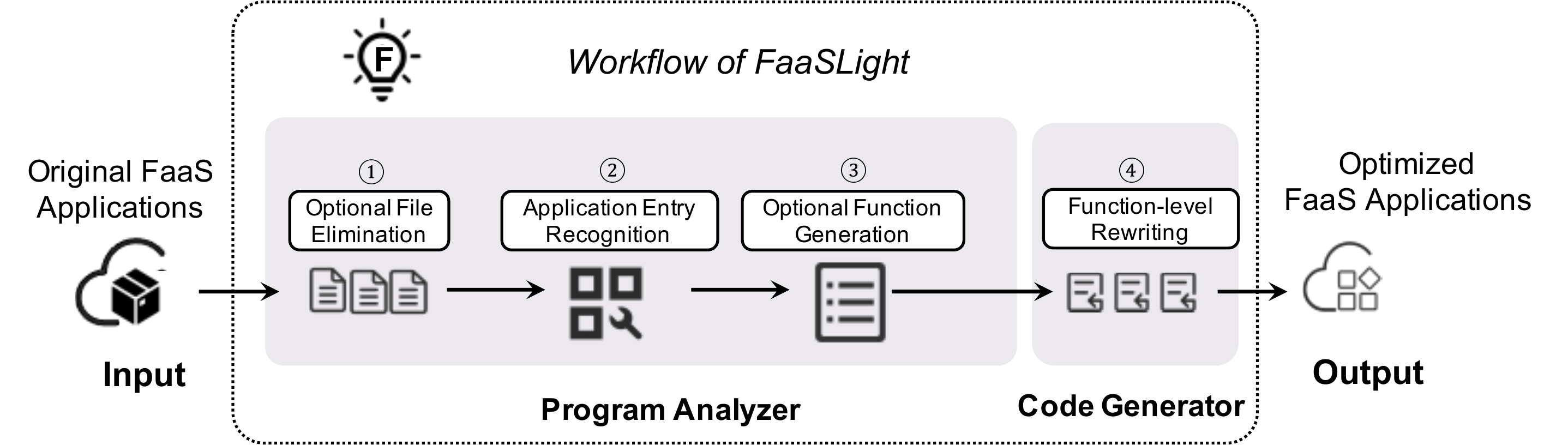}
    \caption{The workflow overview of \toolName.}
    \label{fig:overview}
\end{figure*}

 Fig.~\ref{fig:overview} shows the detailed architecture of \toolName, which contains two parts (i.e., \textbf{Program Analyzer} for optional function identification, and \textbf{Code Generator} for optional function elimination). More specifically, \toolName has four components (i.e., \textcircled{1} to \textcircled{4} in the figure).
Given a FaaS application, \toolName constructs the call graph to generate the final set of optional functions by using \textbf{Program Analyzer} part (Section~\ref{programanalyzer}). Then, 
based on the optional function set, \textbf{Code Generator} part generates the optimized FaaS application by separating optional functions from the application and designing a rewriting approach for optional functions. This rewriting approach can fetch and execute the required optional functions in an on-demand loading manner (Section~\ref{programgenerator}).

\subsection{Program Analyzer}\label{programanalyzer}

\emph{Program Analyzer} part is responsible for obtaining the final optional functions for the FaaS application. It has three components, including \textit{\textcircled{1} Optional File Elimination}, \textit{\textcircled{2} Application Entry Recognition}, and \textit{\textcircled{3} Optional Function Generation}. \textit{\textcircled{1} Optional File Elimination} is responsible for pre-processing the given FaaS application to remove unneeded files, such as log files. Then, \textit{\textcircled{2} Application Entry Recognition} find entries of the given FaaS application. \textit{\textcircled{3} Optional Function Generation} generates the final set of optional functions. Specific details of the three components are as follows.



\textit{\textcircled{1} Optional File Elimination.} It removes files that are not indispensable (i.e., files that are not used when the application is executed) to get a simplified FaaS application. According to the actual development process, \toolName eliminates four types of optional files for FaaS applications. (1) Files related to the virtual environment of local development. Developers may package some local files that are not related to the application functionality. For example, ``pip'' and ``setuptools'' directories may be included in the Python  application; (2) The compiled files (e.g., in  ``pyc'' or ``pyi'' format). These files are generated when developers test their functionalities locally, increasing the FaaS application size; (3) The information-related directories in used general libraries. For example, the ``dist-info'' directory only describes additional information about libraries; (4) Test cases related files in used general libraries. For example, functionalities of the ``tests'' directory in \textit{NumPy} library are not used by developers at all. Deleting these four types of files also can decrease the code analysis complexity of the FaaS application later.

\textit{\textcircled{2} Application Entry Recognition.} Generally, there are two types of functions that act as the entry of a FaaS application, including serverless functions and module initialization functions. In addition, considering the programming features of different languages, there may be language-specific functions that can play this role.
The goal of \textit{Application Entry Recognition} component is to identify these types of entries. Next, we introduce the recognition of each type.

\emph{Serverless functions:} Serverless functions are the interfaces exposed to the Internet. Since FaaS applications are executed in an event-driven fashion, the key to identifying serverless functions is to identify event handlers. We use three identification strategies. First, it is to find the configuration file (e.g., described in ``.yml'' or ``.yaml'' files), which is commonly used to configure the resource and permission of cloud applications~\cite{WenServerless21}. In such a file, the relationship between serverless functions and the corresponding events can be found, further determining the name of the used serverless functions. Second, when such a configuration file does not exist in the deployment package of the FaaS application, the component analyzes the source code to find all function definitions and match the specific parameters' format. For example, on AWS Lambda, input fields of serverless functions are generally filled with ``\texttt{event}'' and ``\texttt{context}''~\cite{AWSLambdahandler}, while Google Cloud Functions-based serverless functions are filled with ``\texttt{request}''~\cite{Googlehandler}. 
Third, if previous strategies do not work, the component provides an external interface to allow developers to define the name of serverless functions. Through these strategies, entry points about serverless functions can be determined for the given FaaS application. Note that there can be optional techniques for event handler identification~\cite{jensen2015stateless, adamsen2018practical, zhao2010Event, kupoluyi2021muzeel}. However, these techniques cannot be directly deployed, since they need to obtain a large number of test cases and use these cases to dynamically execute or interact with the applications, leading to too huge runtime overhead and low efficiency to be feasible in \toolName.

\emph{Module initialization functions:} Module initialization functions are functions that the program needs to first call to initialize the loaded module before loading a module. \toolName detects the module initialization functions through a list generated by offline profiling. The specific process of profiling is explained as follows. We first insert probes (such as the ``\texttt{print}'' statement for the Python program and the ``\texttt{log}'' statement for the JavaScript program) to the entry of all library functions. The probes will record which functions are invoked during the runtime. Then we load common dependency libraries (e.g., \textit{Numpy} for Python and \textit{request} for JavaScript) offline and get which functions are invoked during the initialization stage through the logs generated by the probes.

\emph{Language-specific functions:} There are  additional functions that can act as the application entry according to the characteristics of programming languages. We take Python and JavaScript, the two most widely-adopted programming languages for developing FaaS applications, as examples. For Python-based FaaS applications, magic functions are common entries~\cite{magicfunction}. They are invoked during the use of overloaded operators, and the execution process is automatically done without explicitly specifying the name of the magic function. Therefore,  it is difficult to identify the calling information about such automatic execution behaviors by statically analyzing code. Since magic functions are wrapped in double underscores like ``\_\_xx\_\_'', we recognize functions in such a format as magic functions. 
For JavaScript-based FaaS applications, asynchronous callback functions~\cite{JavaScriptasync} are commonly used entries. They are passed to another function as a parameter, so it is challenging to capture the function-calling information due to asynchronous handling. To tackle this problem, we analyze parameter names to match existing function definitions and then establish the relationships with callback functions. Functions that satisfy this criterion are recognized as asynchronous callback functions.

\textit{\textcircled{3} Optional Function Generation.} It accepts the entries identified by \textcircled{2}, and outputs a set of optional functions. The idea of \textit{Optional Function Generation} is straightforward. It traverses the call graph and gets all functions that are reachable from the entries in the indispensable function set. Then it outputs the functions that are not in the indispensable function set as optional functions.

In our work, we perform a whole program analysis, including serverless function code and the imported third-party libraries, to build call graphs. The specific detail is as follows. We adopt a similar idea of Class Hierarchy Analysis (CHA)~\cite{dean1995optimization,bruce2020jshrink} to identify call targets of functions. CHA is a rapid analysis technique that conservatively estimates possible calling relationships. Moreover, we use context-insensitive and flow-insensitive analyses. Such a design is enough to process serverless functions in flexible practicality and scalability. In our study, the component identifies all related called functions in the definition scope of a given function and considers them as potential call targets. However, since code features are complex and variable, the engineering implementation is not easy and may face additional challenges, e.g., alias recognition and nested functions. These challenges are solved by designing specific identification methods. For example, we analyze the parent code of the nested function to establish their mappings. Finally, the call graph is generated to capture the call relationships among functions, including different types of entries.

\subsection{Code Generator}\label{programgenerator}

\emph{Code Generator} part is the key step in optimizing the code of the FaaS application. It can achieve reduction and on-demand loading of optional code. The input of this part is the set of optional functions, and its output is to generate the optimized FaaS application. \emph{Code Generator} part leverages the \textit{Function-level rewriting} component to achieve its goal.


\textit{\textcircled{4} Function-level Rewriting.} It is to rewrite optional functions from the FaaS application. Two main operations are accomplished: separation and rewriting, which are performed simultaneously. The separation operation separates the optional functions from the application. The workflow of the separation operation is as follows. First, for the optional function to be processed, \toolName saves the whole function definition and the body content into the corresponding value of ``key-value'' pairs in string format. Note that the key of ``key-value'' pairs is the representation of function location and function name. Then, the optional function is transferred as the corresponding function definition with an empty code body. In the rewriting operation, the empty code body is filled with our custom execution code, which has much fewer lines of code (e.g., 2 lines) than their original code. Through such a concise way, the loaded code size of the FaaS application is reduced. An example of the optional function rewriting is shown in Listing~\ref{lst:function-example}, whose code body has 23 lines that do not contain comments. This optional function can be transformed into 2 lines of code shown in Listing~\ref{lst:rewrite-example}. Our custom execution code is to execute the ``\textit{rewrite\_template}'' method from the ``\textit{custom\_functemplate}'' module. Note that if the optional function to be processed has the parent function, and this parent function is also an optional function, the current optional function will not be rewritten, and later \toolName will directly rewrite its parent function. Such a design can reduce the number of rewritings. After handling all optional functions, the content of ``key-value'' pairs is generated and compressed into a global lightweight file through a compression strategy (e.g., ``gzip''). Finally, all optional functions are separated out and also rewritten to finish. The lightweight file is saved in the deployment package of the FaaS application.

\begin{lstlisting}[language={Python}, caption={An example of the original function. (\textit{pandas/compat/pickle\_compat/})}, label={lst:function-example}]
def load_reduce(self):
    stack = self.stack
    args = stack.pop()
    func = stack[-1]
    if len(args) and type(args[0]) is type:
        n = args[0].__name__  # noqa
    try:
        stack[-1] = func(*args)
        return
    except TypeError as err:
        # If we have a deprecated function,
        # try to replace and try again.
        msg = "_reconstruct: First argument must be a sub-type of ndarray"
        if msg in str(err):
            try:
                cls = args[0]
                stack[-1] = object.__new__(cls)
                return
            except TypeError:
                pass
        elif args and issubclass(args[0], BaseOffset):
            # TypeError: object.__new__(Day) is not safe, use Day.__new__()
            cls = args[0]
            stack[-1] = cls.__new__(*args)
            return
        raise
\end{lstlisting}

\begin{lstlisting}[language={Python}, caption={An example of the rewritten function.}, label={lst:rewrite-example}]
def load_reduce(self):
    import custom_functemplate
    return custom_functemplate.rewrite_template("pandas.compat.pickle_compat.load_reduce", "load_reduce(self)", {"BaseOffset": BaseOffset, "self": self}, 1)
\end{lstlisting}

When executing the optimized FaaS application, some optional functions are required. The ``\textit{rewrite\_template}'' method in the optional function first checks whether the global lightweight file is loaded. If it does not exist, the ``\textit{rewrite\_template}'' method can read this file into memory and fetch the required code in the string format. If it exists, the required code can be directly fetched from memory to execute. However, the code execution of fetched optional functions may depend on some necessary variables due to the inconsistent execution context. These variables are external functionalities used in this function, and they can be generated through code analysis while rewriting the optional function.

When a FaaS application finishes all components in Fig.~\ref{fig:overview}, it becomes the optimized FaaS application with the necessary code.

\subsection{Implementation}


We have fully-fledged implemented \toolName with two execution runtime supports, one for FaaS applications written in Python and the other for ones written in JavaScript. Note that the design principles behind these two supports are inherently general according to what is presented in the preceding sections, requiring only some engineering efforts according to language-specific settings. However, these efforts are implemented inside the tool itself, while completely transparent to FaaS developers. In practice, the developer needs to only submit their FaaS application code to \toolName via the tool interface, without introducing any additional efforts. For every single FaaS application, \toolName can generate the corresponding optimized version through four components. In our approach, the main analysis builds on the foundation and improvements of CHA~\cite{dean1995optimization,bruce2020jshrink}, \emph{ast}~\cite{astforPython}, \emph{astroid}~\cite{astroid}, and \emph{uglify-js}~\cite{uglifyjs} libraries. 





%% file: section/evaluation.tex
\section{Evaluation}\label{sec:evaluation}

We evaluate the effectiveness of \toolName on 15 real-world FaaS applications used in Section~\ref{sec:measurement}. We aim to evaluate \toolName by answering the following research questions.

\textbf{RQ1 (Code reduction):} \textit{How much can \toolName reduce the size of FaaS applications?} 


\textbf{RQ2 (Cold performance):} \textit{How much can \toolName speed up cold starts of FaaS applications?} 


\textbf{RQ3 (Warm performance):} \textit{How does \toolName affect the performance of warm starts of FaaS applications?} 


\textbf{RQ4 (Overhead analysis):} \textit{What is the performance overhead introduced by the on-demand loading mechanism of \toolName?} 


\textbf{RQ5 (Comparison):} \textit{How does \toolName perform compared with state-of-the-art methods?}

In addition, we evaluate the generalizability of our approach (i.e., loading only indispensable code) on performance optimization of FaaS applications. We aim to answer the research question about RQ6.


\textbf{RQ6 (Generalizability):} \textit{Can our approach be generalized to FaaS applications written in other languages or executed on other serverless platforms?}

\subsection{Evaluation Settings} 
We evaluate \toolName with the 15 real-world FaaS applications in Table~\ref{tab:datainfo} to answer the research questions of RQ1 - RQ5. \toolName runs on a server with Intel Xeon (R) 4 cores and 24GiB main memory. The system of this server is Ubuntu 18.04.4 LTS. 
The tested FaaS applications are originally executed on AWS Lambda, which is the most popular and widely used serverless platform~\cite{AWSLambdaTop1,AWSLambdaTop2}. In our study, original FaaS applications are denoted as \emph{before} applications. FaaS applications processed by \textit{Optional File Elimination} component of \toolName are denoted as \emph{after1} applications, and the final optimized FaaS applications are denoted as \emph{after2} applications. 

We run experiments on each of the 15 tested FaaS applications 20 times to collect performance metrics (e.g., latency and memory usage) and use \textit{Mann Whitney U-test}~\cite{pham2020problems,MannWhitneyUtest} (which is suitable for the small sample size and does not require normality) to measure the statistical significance. When comparing two sets of performance results for \textit{after2} and \textit{before} applications, the null hypothesis is that the performance of \textit{after2} set is similar to \textit{before} set. The threshold of statistical significance is set as p-value < 0.05. We further compute the effect size as the Cohen's \textit{d}~\cite{effectsize}, to check if the difference has a meaningful effect. \textit{d} is between 0 and 2, where 0.2 indicates a small effect, 0.5 a medium effect, and 0.8 a large effect~\cite{effectsize}.

%% file: section/result.tex
\begin{figure*}[!thb]
	\centering
    \includegraphics[width=0.99\textwidth]{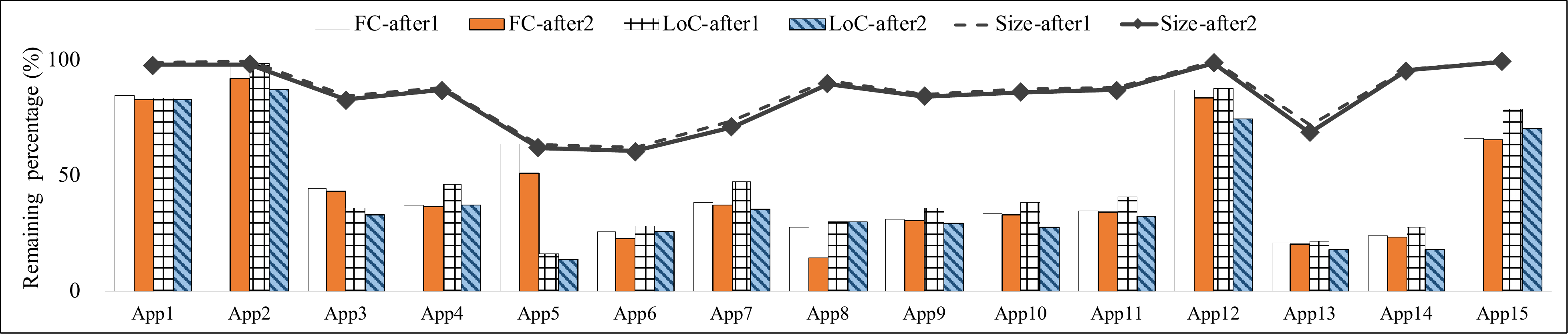}
    \caption{The change of statistical values (the optimized application as a percentage of the original application).}
    \label{fig:app_info_slim}
\end{figure*}

\subsection{RQ1: Code Reduction}



To explore how \toolName reduces the application code size and hence code loading time, we compare the application package size (\emph{Size}), the number of functions (\emph{FC}), and lines of code (\emph{LoC}) for \textit{before}, \textit{after1}, and \textit{after2} applications, respectively. The percentage of these values for \textit{after1} and \textit{after2} applications compared to the \textit{before} application is shown in Fig.~\ref{fig:app_info_slim}. Specifically, in the best case, \emph{Size} of \textit{after2} applications becomes 60.84\% of \textit{before} applications, reducing 39.16\% optional content. On average, \toolName makes the size of \textit{before} applications decrease to their 84.82\%, reducing 15.18\% \emph{Size}. For \emph{FC} and \emph{LoC} of \textit{after2} applications, \toolName reduces by 55.06\% and 58.75\% of the original \emph{FC} and \emph{LoC}, respectively, on average.


From Fig.~\ref{fig:app_info_slim}, we also find that the reduction of statistical values is mainly in \textit{after1} applications, which means that \textit{Optional File Elimination} component can effectively decrease a large part of optional files. Based on \textit{after1} versions, \toolName leverages \textit{Code Generator} part to merge and simplify optional functions that are loaded, i.e., rewriting optional functions, to further decrease statistical values to get the \textit{after2} version. In addition, when executing the optimized FaaS applications, on average, \toolName callbacks only 10 optional functions according to given input cases on demand.


\finding{On average, \toolName reduces the application size by 15.18\%, the number of functions by 55.06\%, and the number of code lines by 58.75\%.}

\subsection{RQ2: Cold Performance}

\begin{table*}[ht]
  \footnotesize
 \centering
 \caption{The performance results of Python-based FaaS applications in cold starts on AWS Lambda. }
 \label{tab:result_cold}
   \begin{tabular}{l|l|l|l|l|l}
   \hline 
    \rowcolor{gray!50}\textbf{App ID}&\textbf{Version}&\tabincell{c}{\textbf{Preparation phase latency} \\\textbf{(ms)}}&\tabincell{c}{\textbf{Loading phase latency }\\\textbf{(ms)}}&\tabincell{c}{\textbf{Runtime memory} \\\textbf{(MB)}}&\tabincell{c}{\textbf{Total response latency} \\ \textbf{(ms)}}  \cr

    \hline

    \multirow{3}*{\tabincell{l}{App1}} & before & 1420.43 & 760.95  & 93 & 4011.32\\ 
     & after1    & 1279.10 (\textbf{-- 9.95\%})  & 748.53   & 93 & 3792.86 (\textbf{-- 5.45\%}) \\
     & after2   & 1270.38 (\textbf{-- 10.56\%}) & 701.82 (\textbf{-- 7.77\%})   & 90 (\textbf{-- 3.23\%}) & 3565.94 (\textbf{-- 11.10\%})\\

    \hline
   \multirow{3}*{\tabincell{l}{App2}} & before  & 1463.63 & 873.00 & 68 & 2468.94 \\
    & after1  & 1296.08 (\textbf{-- 11.45\%})  & 870.86  &  68  & 2316.69 (\textbf{-- 6.17\%})  \\
    & after2  &  1279.88 (\textbf{-- 12.55\%})  & 183.77 (\textbf{-- 78.95\%})  & 43 (\textbf{-- 36.76\%}) & 1602.61 (\textbf{-- 35.09\%}) \\

    \hline
   \multirow{3}*{\tabincell{l}{App3}} & before  & 1532.66 & 1794.47 &  115 & 3522.86\cr
    & after1   & 1343.44 (\textbf{-- 12.35\%})  & 1782.62 &   115  & 3323.86 (\textbf{-- 5.65\%})  \\
   & after2   &  1232.81 (\textbf{-- 19.56\%})  & 1708.84 (\textbf{-- 4.77\%})   & 113 (\textbf{-- 1.74\%}) & 3131.91 (\textbf{-- 11.10\%}) \\

    \hline

   \multirow{3}*{\tabincell{l}{App4}} & before & 2011.29 & 2352.07 & 142 & 4593.43\\ 
   & after1   & 1778.86 (\textbf{-- 11.56\%}) & 2235.33 (\textbf{-- 4.96\%}) &  141 & 4163.48 (\textbf{-- 9.36\%}) \\
    & after2  & 1768.07 (\textbf{-- 12.09\%})  & 2033.82 (\textbf{-- 13.53\%})  & 140 (\textbf{--1.41\%})  & 4004.10 (\textbf{-- 12.83\%})\\
    
    \hline
   \multirow{3}*{\tabincell{l}{App5}} & before  & 1279.98 & 488.81 & 62 & 2689.65\\ 
    & after1  &  1249.72  & 451.97 (\textbf{-- 7.54\%})&   61 & 2601.61 (\textbf{-- 3.27\%}) \\
   & after2 &  1272.52  & 435.84 (\textbf{-- 10.84\%}) &  60 (\textbf{-- 3.23\%}) & 2511.24 (\textbf{-- 6.63\%}) \\

   \hline
   \multirow{3}*{\tabincell{l}{App6}} & before &1540.03 & 1959.92 & 125 & 3660.67 \\ 
    & after1  & 1398.48 (\textbf{-- 9.19\%}) & 1774.72 (\textbf{-- 9.45\%}) & 115 (\textbf{-- 8.00\%}) & 3346.29 (\textbf{-- 8.59\%}) \\
     & after2  & 1340.51 (\textbf{-- 12.96\%}) & 1536.38 (\textbf{-- 21.61\%})  & 107 (\textbf{-- 14.40\%}) & 3054.51 (\textbf{-- 16.56\%}) \\
    \hline 
   
   \multirow{3}*{\tabincell{l}{App7}} & before  & 2312.04 & 4580.31 &  228 & 7165.54\\ 
    & after1  & 2185.66 (\textbf{-- 5.47\%}) & 4217.68 (\textbf{-- 7.92\%}) & 206 (\textbf{-- 9.65\%}) & 6770.75 (\textbf{-- 5.51\%})\\
    & after2  & 2177.49 (\textbf{-- 5.82\%}) & 1408.62 (\textbf{-- 69.27\%})  & 130 (\textbf{-- 42.98\%}) & 4152.73 (\textbf{-- 42.05\%})\\
   
   \hline

   \multirow{3}*{\tabincell{l}{App8}} & before  & 1739.22 & 887.81 &  102 & 2741.02 \\
    & after1  & 1642.62 (\textbf{-- 5.55\%})  & 790.91 (\textbf{-- 10.92\%})  & 98 (\textbf{-- 3.92\%}) & 2562.48 (\textbf{-- 6.51\%}) \\
     & after2  & 1620.82 (\textbf{-- 6.81\%})  & 188.26 (\textbf{-- 78.80\%})  & 42 (\textbf{-- 58.82\%}) & 1951.16 (\textbf{-- 28.82\%}) \\

   \hline

   \multirow{3}*{\tabincell{l}{App9\\train\\}} & before  & 2741.06 & 3985.42 &  230 & 9035.39\\ 
    & after1  & 2140.74 (\textbf{-- 21.90\%}) & 3790.63 \textbf{-- 4.89\%}) &   229   & 8218.25 (\textbf{-- 9.04\%}) \\
    & after2  & 2108.48 (\textbf{-- 23.08\%}) & 3135.82 (\textbf{-- 21.32\%})  & 216 (\textbf{-- 6.09\%}) & 7470.49 (\textbf{-- 17.32\%})\\
   
   \hline
   
    \multirow{3}*{\tabincell{l}{App9\\predict\\}} & before  & 2700.32 & 3828.55 &  230 & 8291.80\\ 
    & after1  &  2188.82 (\textbf{-- 18.94\%}) & 3689.76 (\textbf{-- 3.63\%}) &  229   & 7712.55 (\textbf{-- 6.99\%}) \\
    & after2  & 1994.47 (\textbf{-- 26.14\%}) & 3141.81 (\textbf{-- 17.94\%}) &  215 (\textbf{-- 6.09\%}) & 7071.03 (\textbf{-- 14.72\%})\\

   \hline

   \multirow{3}*{\tabincell{l}{App10}} & before  & 2365.90 & 2391.77 &  159 & 4961.16\\ 
    & after1 & 2081.80 (\textbf{-- 12.01\%}) & 2272.51 (\textbf{-- 4.99\%})  & 158 & 4494.80 (\textbf{-- 9.40\%}) \\
   & after2  & 1914.79 (\textbf{-- 19.07\%})  & 1895.94 (\textbf{-- 20.73\%})  & 148 (\textbf{-- 6.92\%}) & 4035.48 (\textbf{-- 18.66\%})\\
   
   \hline

   \multirow{3}*{\tabincell{l}{App11}} & before &2018.32 & 3384.63  & 182 &5551.03 \\ 
    & after1  & 1943.81 (\textbf{-- 3.69\%}) & 3308.98 (\textbf{-- 2.23\%})  & 181  & 5407.95 (\textbf{-- 2.58\%}) \\
    & after2  & 1949.72 (\textbf{-- 3.40\%})  & 1722.93 (\textbf{-- 49.10\%})  & 141 (\textbf{-- 22.53\%}) & 3934.31 (\textbf{-- 29.12\%})\\

   \hline

   \rowcolor{gray!1}\multirow{3}*{\tabincell{l}{App12\\(Docker)\\}} & before  & 1266.90 & 6966.72  & 410 & 8442.14\\ 
    & after1  & 1255.25 & 6901.35  & 410  & 8281.06 (\textbf{-- 1.91\%})  \\
   & after2   & 1279.19 & 6036.30 (\textbf{-- 13.36\%})    & 397 (\textbf{-- 3.17\%}) & 7448.55 (\textbf{-- 11.77\%})\\
   
    \hline

    \rowcolor{gray!1}\multirow{3}*{\tabincell{l}{App13}} & before  & 1441.96 & 701.41  & 78 & 2304.29 \\
     & after1  & 1252.35 (\textbf{-- 13.15\%})  & 646.34 (\textbf{-- 7.85\%})  & 74 (\textbf{-- 5.13\%}) & 2052.75 (\textbf{-- 10.92\%}) \\
     & after2 & 1241.88 (\textbf{-- 13.88})  & 571.64 (\textbf{-- 18.50\%})  & 69 (\textbf{-- 11.54\%}) & 1961.71 (\textbf{-- 14.87\%}) \\
     
   \hline

   \multirow{3}*{\tabincell{l}{App14}} & before  & 2592.21 & 953.36 & 116 & 3678.09 \\ 
   & after1  & 2153.25 (\textbf{-- 16.93\%}) & 938.65 (\textbf{-- 1.54\%})   & 116 & 3195.03 (\textbf{-- 13.13\%}) \\
    & after2  & 2032.94 (\textbf{-- 21.58\%})  & 830.00 (\textbf{-- 12.94\%})   & 114 (\textbf{-- 1.72\%}) & 2980.90 (\textbf{-- 18.96\%})\\

    \hline
   
   \multirow{3}*{\tabincell{l}{App15\\(Docker)\\}} & before & 1368.85 & 6211.63  & 872 & 8062.10\\ 
   & after1   & 1319.33  & 5923.85 (\textbf{-- 4.63\%}) &  872 & 7840.81 (\textbf{-- 2.74\%}) \\
   & after2   & 1307.50   & 4902.64 (\textbf{-- 21.07\%})    & 732 (\textbf{-- 16.06\%})  & 6635.88 (\textbf{-- 17.69\%})\\
  
  \hline 
  \textbf{Max} &  &  \textbf{26.14\%} & \textbf{78.95\%} &  \textbf{58.82\%} & \textbf{42.05\%} \\ 
  \textbf{Mean} &  &  \textbf{11.72\%} & \textbf{28.78\%} & \textbf{14.79\%} & \textbf{19.21\%}\\

   \hline
   \end{tabular}
\end{table*}


\begin{figure}[!thb]
	\centering
    \includegraphics[width=0.6\textwidth]{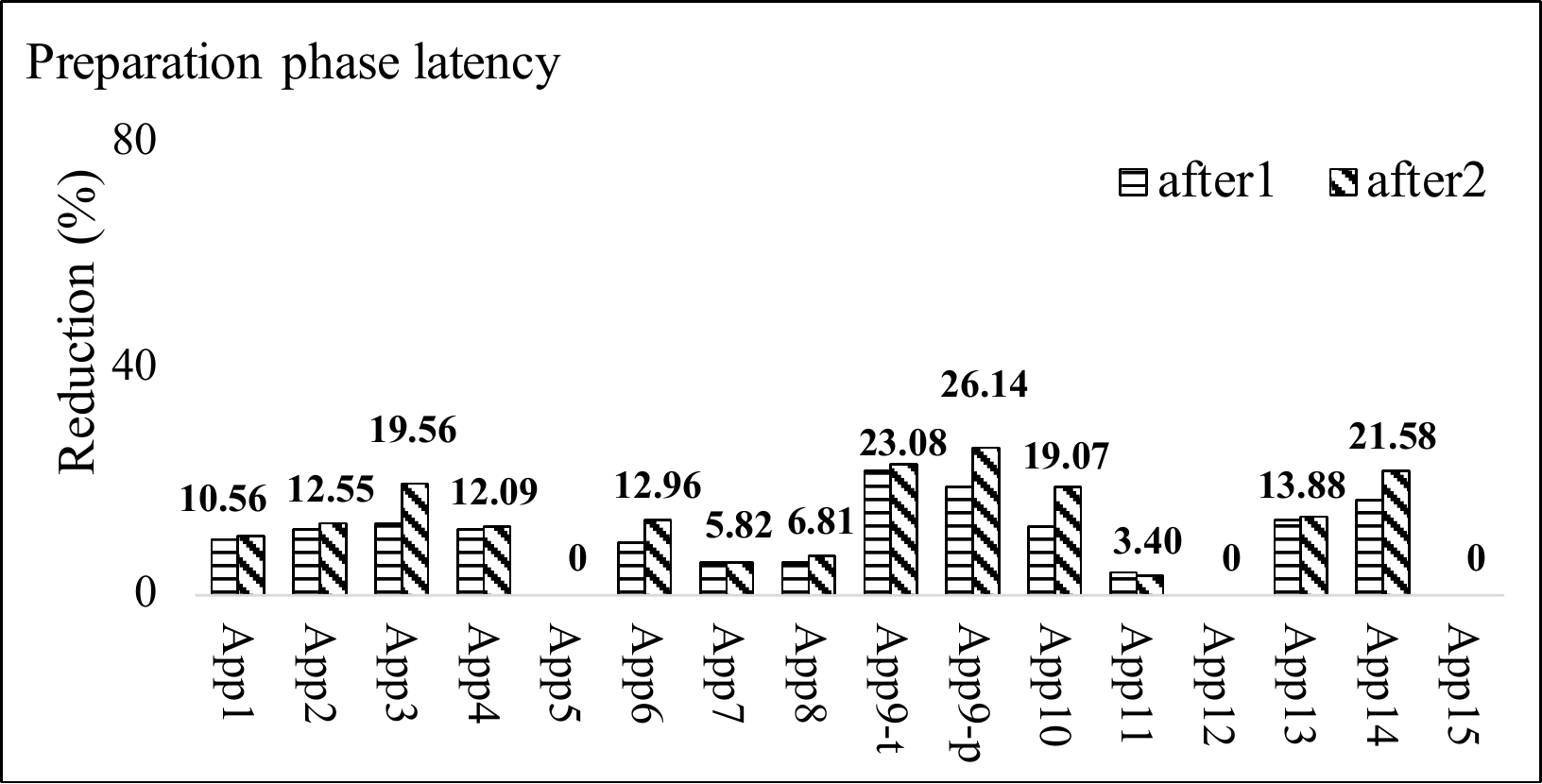}
    \caption{The reduction percentage of the preparation phase latency for \textit{after1} and \textit{after2} applications in cold starts.}
    \label{fig:PrepareLatencyReductionPrecentage}
\end{figure}

\begin{figure}[!thb]
	\centering
    \includegraphics[width=0.6\textwidth]{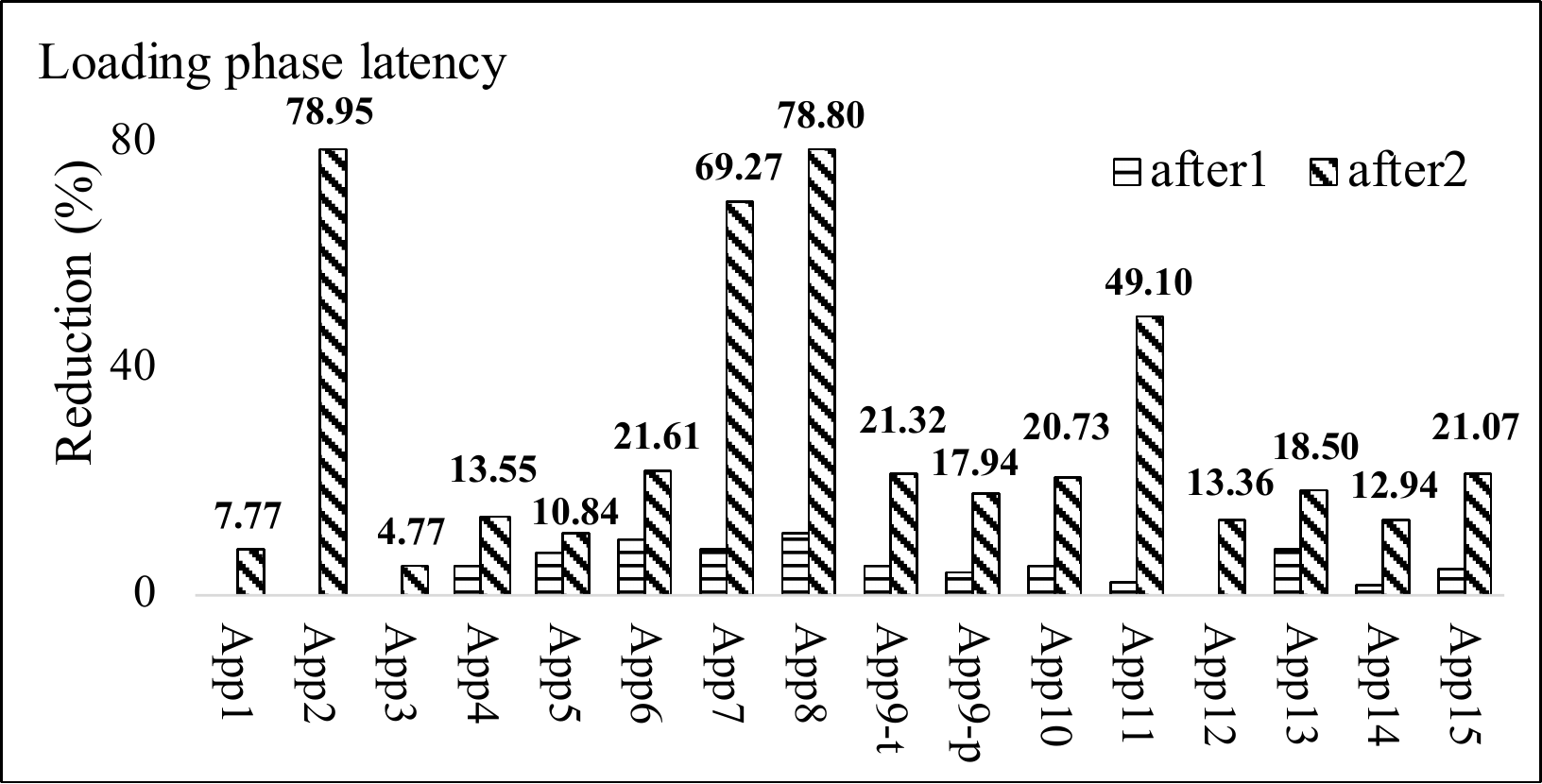}
    \caption{The reduction percentage of the loading phase latency for \textit{after1} and \textit{after2} applications in cold starts.}
    \label{fig:LoadLatencyReductionPrecentage}
\end{figure}



\begin{figure}[!thb]
	\centering
    \includegraphics[width=0.6\textwidth]{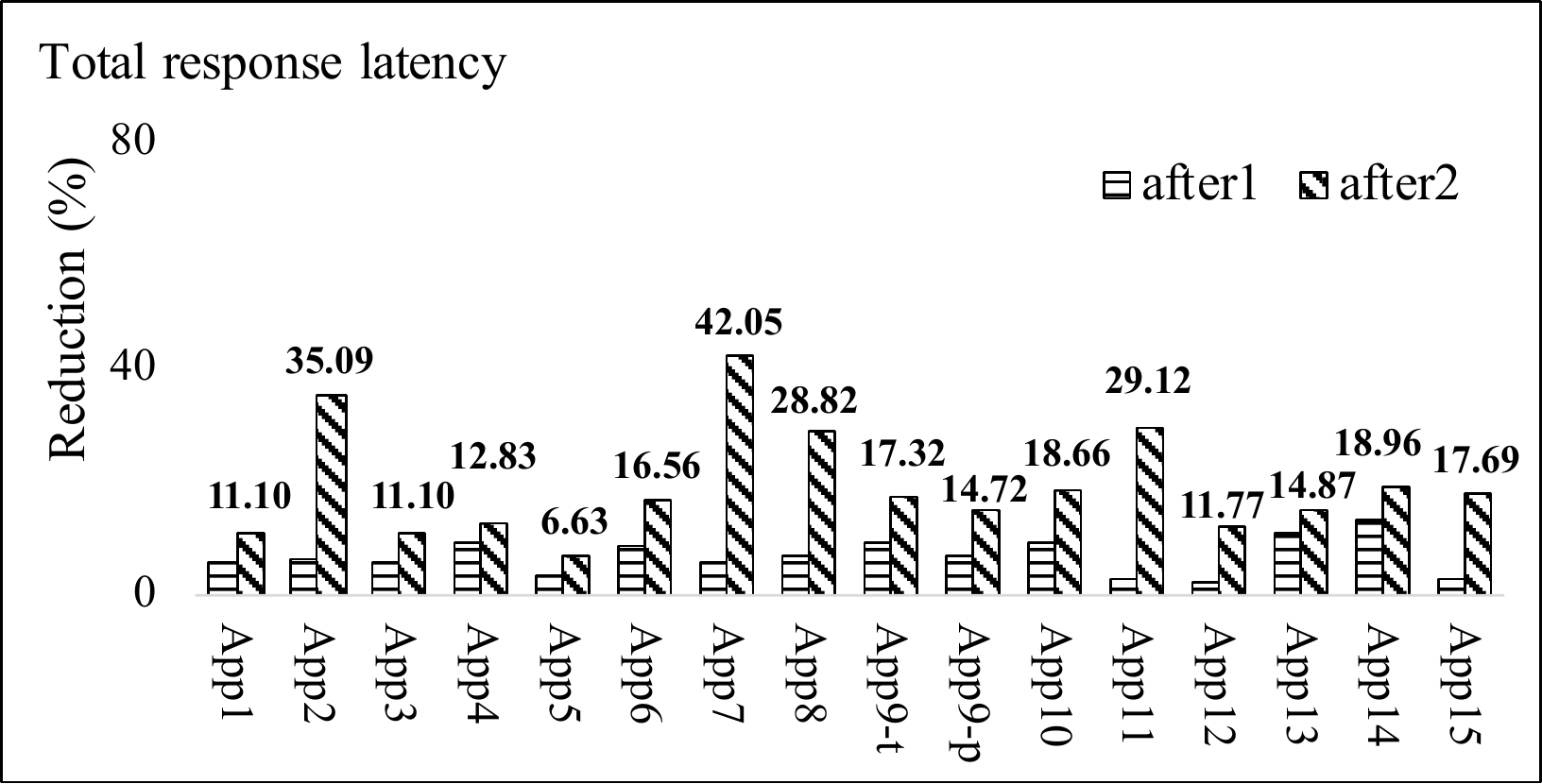}
    \caption{The reduction percentage of the total response latency for \textit{after1} and \textit{after2} applications in cold starts.}
    \label{fig:TotalLatencyReductionPrecentage}
\end{figure}

To answer how \toolName speeds up the performance in cold starts, we compare the preparation phase latency, loading phase latency, total response latency, and runtime memory for \textit{before}, \textit{after1}, and \textit{after2} applications, respectively. The specific explanation of these latencies can be found in Section~\ref{sec:measureresult}. In our approach, the performance improvement of FaaS applications is mainly attributed to two components. The first one is the \textit{Optional File Elimination} component, which reduces optional files. The second one is the \textit{Function-level Rewriting} component, which separates optional code from the loaded application code. \textit{Optional File Elimination} component can reduce the application size, thus decreasing the application transmission time, i.e., the preparation phase latency. \textit{Function-level Rewriting} component can directly decrease the loading code size for FaaS applications, thus reducing the loading phase latency. Table~\ref{tab:result_cold} shows the real latency results for tested FaaS applications. Meanwhile, Figs.~\ref{fig:PrepareLatencyReductionPrecentage}, \ref{fig:LoadLatencyReductionPrecentage}, and \ref{fig:TotalLatencyReductionPrecentage} show reduction percentages of the preparation phase latency, loading phase latency, and total response latency for \textit{after1} and \textit{after2} applications in cold starts, respectively. Specifically, as shown in Fig.~\ref{fig:PrepareLatencyReductionPrecentage}, \toolName reduces 3.40\% to 26.14\% of the preparation phase latency except for App5, App12, and App15. The potential reasons are explained as follows. The application size of App5 is small (i.e., 25.26 MB), so the reduction may not be enough to influence the application transmission latency in the preparation phase. Application code sizes of App12 and App15 are large, but their optional files are few since the application consists mainly of dependency libraries with few local development files. Therefore, the preparation phase latencies of App5, App12, and App15 do not change much. \toolName reduces the preparation phase latency by up to 26.14\% (11.72\% on average).

For the loading phase latency, \textit{Code Generator} part is an effective way to decrease this latency by rewriting optionally loaded functions as ones with only two lines of code. Specifically, \toolName makes the loading phase latency reduced by up to 78.95\%. On average, FaaS applications have 28.78\% performance improvement on the loading phase latency. Especially for App2, App7, and App8, \toolName can reduce more than 60\% loading phase latency due to the simplicity of the tasks.


For the total response latency, the reduction percentage for \textit{after1} and \textit{after2} applications is shown in Fig.~\ref{fig:TotalLatencyReductionPrecentage}. Specifically, for the final \textit{after2} applications, \toolName makes the total response latency reduced by up to 42.05\% (19.21\% on average). However, we also observe that some FaaS applications, e.g., App1, App5, and App12, obtain a small performance improvement. The potential reasons are explained as follows. First, these FaaS applications have a low improvement in the preparation phase latency, as previously mentioned. Second, since the functionalities of dependency libraries are well written, the optional code that can be separated from the FaaS application is small, which makes the reduction in the loading phase latencies limited. To explore the performance improvement effect, we calculate \textit{Mann Whitney U-test} of all measurements about the total response latency between \textit{after2} and \textit{before} applications. Results are shown in Table~\ref{tab:effectsize}, where ``*'' represents that the p-value is less than 0.05. We observe that 14 (14/15 = 93.33\%) optimized FaaS applications have a statistically different performance from their original FaaS applications. In these FaaS applications, eight tested FaaS applications (App1, App2, App4, App6, App7, App8, App9-t, App9-p, and App11) show large effect sizes (>= 0.8), i.e., large performance improvement effect. Five tested FaaS applications (App3, App10, App13, App14, and App15) show medium effect sizes (>= 0.5), i.e., medium performance improvement effect. Moreover, the effect sizes of these five FaaS applications are nearly 0.8, indicating that they have a relatively large performance improvement effect at the medium effect level. Only one tested FaaS application (App5) shows a small effect size (>= 0.2), i.e., a small performance improvement effect. However, its effect size is 0.48, which is nearly 0.5. Similarly, App5 shows a relatively large performance improvement effect at the small effect level. Overall, \toolName can significantly improve the total performance of FaaS applications.


\begin{table*}[ht]
  \begin{threeparttable}
    \caption{Statistical test results of the total response latency between \textit{after2} and \textit{before} applications in cold starts. The symbol ``*'' represents that the p-value is less than 0.05.}
    \label{tab:effectsize}

      \begin{tabular}{c|c|c|c|c|c|c|c|c}
      \hline 
        & App1&App2&App3&App4&App5&App6&App7&App8\cr
        
       \hline
      \textbf{Effect size} & 0.81* &  0.96* & 0.74*& 0.80*&0.48*&0.84*&0.99*&0.87*\\
      
      \hline 
        &App9-t&App9-p&App10&App11&App12&App13&App14&App15\cr
        
       \hline
      \textbf{Effect size} & 0.85*&0.80*&0.79*&0.98*&---&0.79*&0.78*&0.75*\\
      \hline
      \end{tabular}
      
      \end{threeparttable}
     
  \end{table*}

\begin{figure}[!thb]
	\centering
    \includegraphics[width=0.6\textwidth]{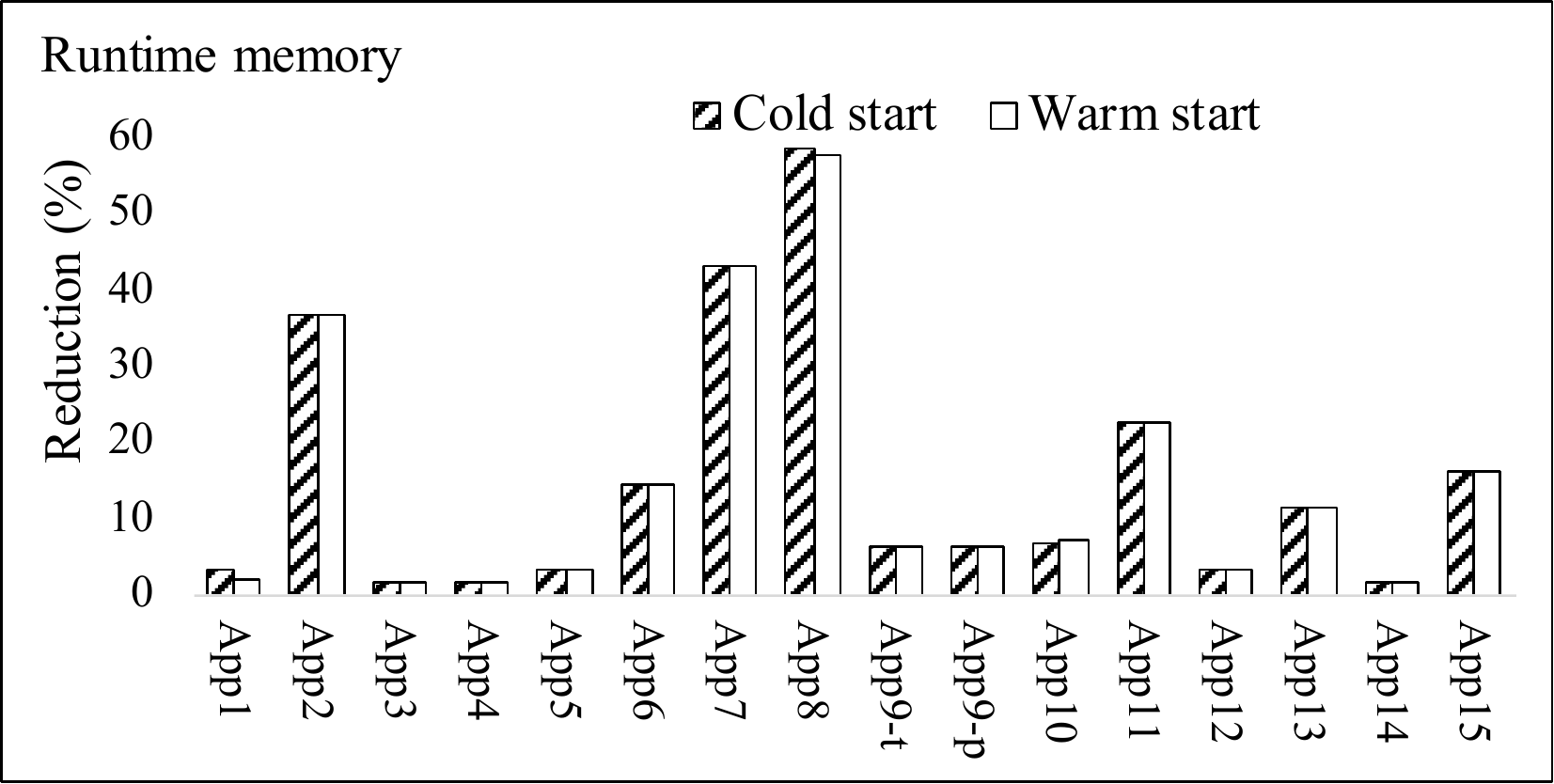}
    \caption{The reduction percentage of the runtime memory for the final \textit{after2} applications in cold and warm starts.}
    \label{fig:memoryusagecoldwarm}
\end{figure}


As shown in Fig.~\ref{fig:app_info_slim}, \textit{after1} applications that apply \textit{Optional File Elimination} component of \toolName can reduce more optional files. However, these optional files are mostly from files unrelated to the application loading code. Directly deleting such files helps FaaS applications decrease the application transmission latency in the preparation phase. The application transmission latency accounts for a small percentage of overhead in the cold-start latency. Therefore, the improvement effect of \textit{after1} applications is limited, shown in Fig.~\ref{fig:LoadLatencyReductionPrecentage} and Fig.~\ref{fig:TotalLatencyReductionPrecentage}. \textit{Program Analyzer} part identifies optional functions, which are major consumers of the application code loading latency. Therefore, as shown in Fig.~\ref{fig:LoadLatencyReductionPrecentage} and Fig.~\ref{fig:TotalLatencyReductionPrecentage}, the performance improvement of \textit{after2} applications is more effective than that of \textit{after1} applications.

\toolName makes the runtime memory reduce by up to 58.82\%  (on average 14.79\%) in cold starts, shown in Fig.~\ref{fig:memoryusagecoldwarm}. In addition, in our study, some tested FaaS applications (e.g., App12 and App15) have a ``big'' application code size, exceeding the normal deployment size limit. They are deployed by the container image. 
The billed duration of this deployment way is the sum of the function execution latency and application code loading latency. Thus, reducing the loading phase latency is beneficial to reduce the developer's billed duration. Results show that \toolName reduces 13.34\% to 20.71\% billed duration of heavy FaaS applications, such as App12 and App15.



\finding{\toolName reduces the preparation phase latency by up to 26.14\% (on average 11.72\%), application code loading latency by up to 78.95\% (on average 28.78\%), total response latency by up to 42.05\% (on average 19.21\%). Moreover, the performance improvement achieved by \toolName is statistically significant for 93.33\% of the studied FaaS applications. As an additional benefit, \toolName decreases the runtime memory by up to 58.82\% (on average 14.79\%).}

\subsection{RQ3: Warm Performance}
\label{sec:rq3}

To explore the effect of \toolName on warm starts of FaaS applications, we compare their scheduling phase latency, total response latency, and runtime memory. Results show that \toolName does not increase the scheduling phase latency and total response latency, meaning that the performance is maintained in the original warm execution performance. We also calculate \textit{Mann Whitney U-test} for all measurements about the total response latency between \textit{after2} and \textit{before} applications. The p-value is all large than 0.05, indicating that optimized applications do not have statistically different performances than original ones in warm starts. We further explore the runtime memory and find that \toolName also reduces the runtime memory by up to 57.84\% (on average 14.74\%) in warm starts shown in Fig.~\ref{fig:memoryusagecoldwarm}. It guides developers to configure lower billing memory.


\finding{\toolName has no observable effect on the performance of FaaS applications in warm starts and reduces the runtime memory by up to 57.84\% (on average 14.74\%).}


\subsection{RQ4: Overhead Analysis}\label{sec:rq4}

\toolName adopts an on-demand loading strategy that fetches optional functions when they are required. This strategy may potentially increase the function execution latency and, therefore, increase its warm-start latency. In Section~\ref{sec:rq3}, we confirm that \toolName does not introduce observable latency to warm starts. This section further studies how \toolName causes runtime overhead and why it does not cause observable delays to warm starts of the FaaS applications. 

We measure the runtime cost introduced by the on-demand loading strategy to answer this research question. We find that the latency overhead of this strategy is about 100 ms, on average, due to reading the lightweight file that saves separated optional functions. Certainly, the reading latency is affected by this file size. When this file saves about 5,000 optional functions, its size is only about 1 MB due to our content compression strategy. Its reading latency is between 110 ms to 150 ms. 


The runtime cost of the on-demand loading strategy does not affect the warm-start latency of FaaS applications because it is a one-time cost if the container or VM of a serverless function is not released. Specifically, when the first optional function in invoked, \toolName reads the lightweight file and loads all optional functions into the memory of the container or VM. As long as the serverless platform does not recycle the function execution environment, optional functions already loaded can be used to assist in subsequent executions of the serverless function. Assume a container instance of a serverless function serves ten requests before it is released. In this case, on-demand loading only happens to the request that first invokes an optional function. This request produces the additional runtime cost, which is added to the function execution latency. However, the other nine requests will not be affected since the memory has loaded all optional functions. Note that the ten requests consist of one cold start and nine warm starts. Although the overhead (100 ms) of the on-demand loading strategy increases the function execution latency, \toolName has made the application code loading latency reduce by 28.78\% on average (about 1,000 ms). Therefore, it is beneficial to trade the one-time 100 ms execution latency for the reduction of the cold-start latency. Moreover, it is worth sacrificing a little execution performance overhead for greater overall performance optimization (on average, 19.21\%) of FaaS applications.

\finding{The on-demand loading strategy of \toolName introduces a small runtime cost (about 100 ms).}

\subsection{RQ5: Comparison}

\begin{figure*}[t]
	\centering
    \includegraphics[width=\textwidth]{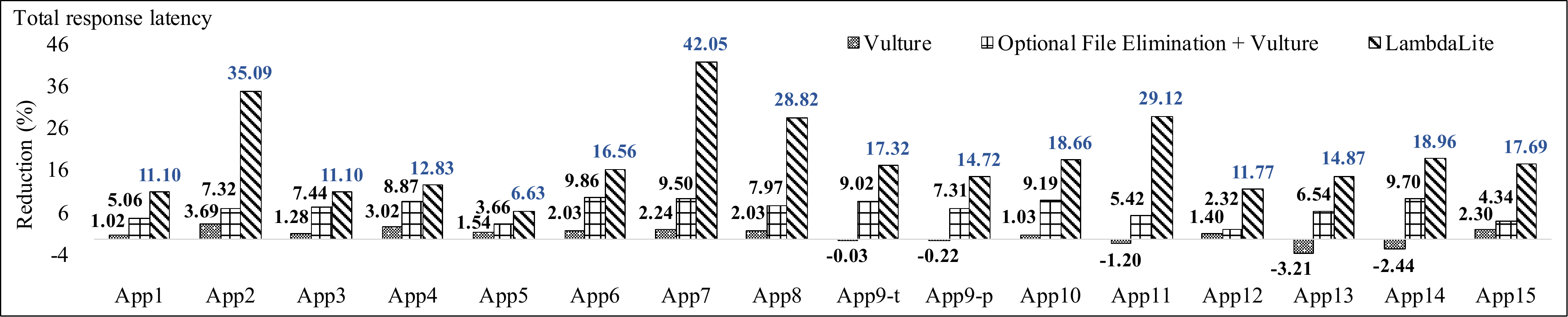}
    \caption{The reduction percentage of the total response latency for different methods in cold starts.}
    \label{fig:CompareTotalLatency}
\end{figure*}



To further demonstrate the effectiveness of \toolName, we compare it with a well-known program analysis tool called \textit{Vulture}~\cite{Vulture}, which can also identify optional code for Python applications and has been widely adopted in the industry~\cite{unusedVulture1,unusedVulture2}. 




\textit{Vulture} identifies the objects that have been defined but not used in all given Python files, and reports them as optional code.
\textit{Vulture} and \toolName are both not related to input cases, focusing on statically analyzing the application code. In our study, we apply \textit{Vulture} to our benchmarks to obtain optional functions. Moreover, we also use a mixed method that combines the functionality of \textit{Optional File Elimination} component and \textit{Vulture}. Optional functions identified by \textit{Vulture} are separated and rewritten by \textit{Code Generator} part of \toolName, in order to be able to load them on demand to ensure the availability of the optimized FaaS application. The results of these methods are shown in Fig.~\ref{fig:CompareTotalLatency}. 

We first compare the performance improvement of \textit{Vulture} method and \toolName. On average, \textit{Vulture} method shows 0.90\% performance improvement on the total response latency, while \toolName can achieve 19.21\% improvement. It illustrates that the latency improvement of \toolName is 21.25$\times$ that of \textit{Vulture}. In addition, in the best case, \textit{Vulture} method obtains 3.69\% performance improvement, while \toolName achieves 42.05\% improvement. Moreover, \textit{Vulture} method does not achieve performance improvement on some FaaS applications, such as App9, App11, App13, and App14. The reason is that the number of optional functions verified by \textit{Vulture} is small (only 400 on average). When the optimized FaaS application needs some optional functions to trigger the on-demand loading, the latency improvement brought by separating optional functions from the application is not enough to compensate for the overhead of reading the file of optional functions. On the contrary, \toolName removes some optional files through \textit{Optional File Elimination} component, and separates on average 5,000 optional functions that are loaded through \textit{Code generator} part. Thus, \toolName shows the effective performance improvement in all tested FaaS applications. 

We further analyze the design principle of \textit{Vulture}. It identifies only the function objects that have been defined but not used in code. Such an analysis lacks a global overview of function usage related to application functionalities. Some functions may be both defined and used in code, but they may be optional for application functionalities. In this situation, \textit{Vulture} misses many functions that may be optional, making the number of separated optional functions small. Therefore, \textit{Vulture} is not effective enough to optimize the total response latency of FaaS applications. On the contrary, \toolName can identify the relevant reachable functions starting from entry points, i.e., serverless functions that represent application functionalities. Functions not related to application functionalities are viewed as optional functions and separated from the FaaS applications.


We also compare the impact of critical parts on the improvement of the total response latency. Compared with the \textit{Vulture} method and the mixed method, the effect of our \textit{Optional File Elimination} component can be analyzed. The mixed method obtains a 7.09\% improvement in reducing the total response latency on average. It illustrates that our \textit{Optional File Elimination} component has a positive impact on performance improvement, speeding up \textit{Vulture} 7.85$\times$. Compared with the mixed method and \toolName, the ability to identify optional functions of loaded code can be analyzed for \textit{Vulture} and our key \textit{Program Analyzer} part. Results show that \toolName improves by 12.12\% on the basis of the mixed method. In this situation, the reduction of the total response latency of \toolName is 2.71$\times$ that of the mixed method. It illustrates that our \textit{Program Analyzer} part is stronger than \textit{Vulture} on the effectiveness of the optional function identification. 
To sum up, critical parts of \toolName have a positive impact on optimizing the total response latency of serverless functions.


\finding{Compared with the state-of-the-art, \toolName achieves a 21.25$\times$ improvement in reducing the average total response latency. 
}

\subsection{RQ6: Generalizability}

For the performance optimization of FaaS applications, we adopt the approach of loading only indispensable code for FaaS applications. To demonstrate the generalizability of the proposed approach, we compare the performance of FaaS applications written in different programming languages (i.e., Python and JavaScript) and executed on different serverless platforms (i.e., AWS Lambda and Google Cloud Functions) before and after optimization, respectively. 

\subsubsection{Language generalizability}\label{sec:rq61}
To demonstrate the language generalizability of \toolName, we further evaluate it on FaaS functions written in JavaScript, which as well as Python are two of the most widely used and popular programming languages in the serverless community~\cite{serverlesscommunitysurvey, serverlesscommunitysurvey1, eismann2021state, eskandani2021wonderless}. To this end, we implement our approach as the JavaScript prototype based on call graph analysis~\cite{antal2018static, javascriptcg}, our customization of program analysis, and optimization of on-demand loading strategy. The source code has been released in our GitHub repository. 

We select the JavaScript-based FaaS applications from the ``Wonderless'' dataset~\cite{eskandani2021wonderless}. This dataset collects applications developed in a serverless fashion from GitHub, and these applications are mainly executed on AWS Lambda. Considering the functionality executability of FaaS applications and the scale representativeness of their application size, finally, we select four FaaS applications written in JavaScript and executed on AWS Lambda. Table~\ref{tab:RQ6JS} shows their specific information, including the application size (\textit{Size}) and the number of lines of code (\textit{LoC}). These FaaS applications are processed by our approach. We compare the related latencies of \textit{before} and \textit{after2} applications in cold starts. The result in Table~\ref{tab:RQ6JS} shows that the preparation phase latency, loading phase latency, and total response latency of \textit{after2} applications are all lower compared with \textit{before} applications. Specifically, the preparation phase latency decreases by 4.27\% to 11.41\%, and our approach reduces the loading phase latency by up to 18.38\%. Finally, the optimized FaaS applications obtain 8.34\% to 10.44\% performance improvement in the total response latency. Overall, these results show that loading only indispensable code can improve the performance of FaaS applications written in JavaScript.

\begin{table*}[t]
  \footnotesize
 \centering
 \caption{The performance result of FaaS applications written in JavaScript and executed in AWS Lambda (cold starts).}
 \label{tab:RQ6JS}
   \begin{tabular}{l|l|l|l|l|l|l|l}
   \hline 
    \rowcolor{gray!50}\textbf{App ID}&\textbf{Name}&\textbf{Version}&\tabincell{c}{\textbf{Size (MB)}}&\tabincell{c}{\textbf{LoC (k)}}&\tabincell{c}{\textbf{Preparation phase } \\\textbf{latency (ms)}}&\tabincell{c}{\textbf{Loading phase}\\\textbf{latency (ms)}}&\tabincell{c}{\textbf{Total response} \\ \textbf{latency (ms)}}  \cr

    \hline

    \multirow{2}*{\tabincell{l}{App16}} & \multirow{2}*{\tabincell{l}{lambda-request~\cite{Test16}}}&before& 4.29&51.32  & 1377.12 & 355.44  & 2368.31\\ 
    & & after2 & 3.38 & 30.80    & 1290.61 (\textbf{-- 6.28\%}) & 303.69 (\textbf{-- 14.56\%})   & 2137.97 (\textbf{-- 9.73\%})\\

    \hline

    \multirow{2}*{\tabincell{l}{App17}} & \multirow{2}*{\tabincell{l}{handlebars-fetch~\cite{Test17}}}&before&  10.21 &125.91  & 1195.14 & 229.29  & 1892.45 \\ 
    & & after2 & 6.67 & 954.19    & 1058.72 (\textbf{-- 11.41\%}) & 187.15 (\textbf{-- 18.38\%})   & 1694.88 (\textbf{-- 10.44\%})\\

    \hline

    \multirow{2}*{\tabincell{l}{App18}} & \multirow{2}*{\tabincell{l}{request-cheerio~\cite{Test18}}}&before&  13.03 &161.42  & 1177.60 & 457.41  & 1853.85 \\ 
    & & after2 & 11.36 & 111.21    & 1084.40 (\textbf{-- 7.91\%}) & 404.42 (\textbf{-- 11.59\%})   & 1699.21 (\textbf{-- 8.34\%})\\

    \hline

    \multirow{2}*{\tabincell{l}{App19}} & \multirow{2}*{\tabincell{l}{serverless-image~\cite{Test19}}}&before&  101.95 &499.43  & 970.98 & 744.49  & 2088.38 \\ 
    & & after2 & 97.21 & 383.52  & 929.47 (\textbf{-- 4.27\%}) & 677.19 (\textbf{-- 9.04\%})   & 1907.12 (\textbf{-- 8.68\%})\\

   \hline
   \end{tabular}
\end{table*}

\subsubsection{Platform generalizability}
To demonstrate the platform generalizability of our approach, we use Google Cloud Functions, another widely adopted serverless platform~\cite{serverlesscommunitysurvey1, AWSLambdaTop1, AWSLambdaTop2, DuASPLOS2020, fouladi2019laptop}, for evaluation.

First, we evaluate \toolName for the aforementioned FaaS applications, i.e., 15 Python applications (i.e., App1 to App15) and 4 JavaScript applications (i.e., App16 to App19), on Google Cloud Functions. These applications were developed for AWS Lambda, and cannot be directly migrated to and executed on  Google Cloud Functions without additional engineering efforts. This is because different serverless platforms may have different requirements for executing the applications. For example, AWS Lambda and Google Cloud Functions have different definition formats of serverless functions~\cite{AWSLambdahandler, Googlehandler}.
To tackle this problem, we manually convert these applications for AWS Lambda to meet the requirements of Google Cloud Functions. To alleviate the potential threat in the implementation of application conversion and migration, we have invited a professional developer, who has five-year practice and experience in developing industry-level FaaS applications, to help review and test our modified code and ensure their correctness. We find that App12 and App15 cannot be migrated and executed, since they exceed the 500 MB limit of the deployment package size that is supported by Google Cloud Functions ~\cite{googleQuotas}. Therefore, we use the remaining 13 applications for evaluation. All the source code of these applications along with the instructions on how we modify the code, have been released on our GitHub.


Table~\ref{tab:RQ6Google} shows the performance of these applications on Google Cloud Functions. The result shows that Google Cloud Functions has a similar cold start problem as AWS Lambda, i.e., the cold-start latency contains the non-negligible application code loading overhead. Thus, reducing loaded code size can help to optimize the application performance. Specifically, on Google Cloud Functions, our approach can reduce the total response latency of serverless functions by up to 32.20\%, with the help of 2.92\% to 14.77\% of preparation phase latency optimization and 8.87\% to 77.15\% of loading phase latency optimization. Overall, these results show that loading only indispensable code can improve the performance of FaaS applications executed on Google Cloud Functions.

To further evaluate the platform generalization, we also collect real-world FaaS applications that were originally designed and executed on Google Cloud Functions. We mine such applications from GitHub using the ``Google Cloud Functions'' keyword. Considering the functionality executability of FaaS applications and the scale representativeness of their application size, we finally select three new  FaaS applications written in Python. We show the details of these applications (i.e., App20, App21, and App22) in Table~\ref{tab:newdataGoogle}. We evaluate \toolName through these applications on both Google Cloud Functions and AWS Lambda. To this end, we follow the previous steps to convert these applications into the format required by AWS Lambda. Table~\ref{tab:RQ6App2022} presents the performance results of these applications. The results show that the effectiveness of \toolName holds for these applications on both Google Could Functions and AWS Lambda, further demonstrating the platform generalization of \toolName. Specifically, on Google Could Functions, \toolName reduces around 12.62\% of the total response latency for these applications; on AWS Lambda, \toolName reduces around 11.37\% of the total response latency. The different performance for different platforms can be explained by the findings in previous studies~\cite{wen2021characterizing, WangATC2018, lee2018evaluation} that different platforms can exhibit different performance results even for the same application.

\begin{table*}[t]
\begin{threeparttable}
  \footnotesize
 \centering
 \caption{The cold start result of App1 to App19 (originally developed for AWS Lambda) on Google Cloud Functions. App12 and App15 are excluded because they exceed the 500 MB uncompressed size limit of Google Cloud Functions.}
 \label{tab:RQ6Google}
   \begin{tabular}{l|l|l|l|l}
   \hline 
    \rowcolor{gray!50}\textbf{App ID}&\textbf{Version}&\textbf{Preparation phase latency (ms)}&\textbf{Loading phase latency (ms)}&\textbf{Total response latency (ms)} \cr

    \hline
    \multirow{2}*{\tabincell{l}{App1}} &before  & 3090.93 & 1410.13  & 6006.80 \\ 
    &  after2     & 2634.28 (\textbf{-- 14.77\%}) & 1258.78 (\textbf{-- 10.73\%})   & 5252.11 (\textbf{-- 12.56\%})\\

    \hline
    \multirow{2}*{\tabincell{l}{App2}} &before  & 2357.36 & 1172.75  & 3859.45 \\ 
    &  after2     & 2200.20 (\textbf{-- 6.67\%}) & 267.97 (\textbf{-- 77.15\%})   & 2764.36 (\textbf{-- 28.37\%})\\

    \hline
    \multirow{2}*{\tabincell{l}{App3}} &before  & 2679.33 & 2921.13  & 5957.00 \\ 
    &  after2     & 2348.42 (\textbf{-- 12.35\%}) & 2594.37 (\textbf{-- 11.19\%})   & 5312.04 (\textbf{-- 10.83\%})\\

    \hline
    \multirow{2}*{\tabincell{l}{App4}} & before  & 2546.22 & 3990.18  & 6923.60 \\ 
    &  after2     & 2389.02 (\textbf{-- 6.17\%}) & 3503.52 (\textbf{-- 12.20\%})   & 6222.04 (\textbf{-- 10.13\%})\\

    \hline
    \multirow{2}*{\tabincell{l}{App5}} & before  & 2520.70 & 667.05  & 5062.62 \\ 
    & after2     & 2447.07 (\textbf{-- 2.92\%}) & 607.86 (\textbf{-- 8.87\%})   & 4519.87 (\textbf{-- 10.72\%})\\

     \hline
    \multirow{2}*{\tabincell{l}{App6}} & before  & 2664.78 & 2925.26  & 5904.28 \\ 
    & after2     & 2351.88 (\textbf{-- 11.74\%}) & 2417.19 (\textbf{-- 17.37\%})   & 5124.88 (\textbf{-- 13.20\%})\\

    \hline
    \multirow{2}*{\tabincell{l}{App7}} & before  & 2700.94 & 4444.32  & 7818.41 \\ 
    &  after2     & 2490.25 (\textbf{-- 7.80\%}) & 2203.49 (\textbf{-- 50.42\%})   & 5301.07 (\textbf{-- 32.20\%})\\

    \hline
    \multirow{2}*{\tabincell{l}{App8}} & before  & 2727.69 & 1289.20  & 4369.73 \\ 
    & after2     & 2548.21 (\textbf{-- 6.58\%}) & 326.21 (\textbf{-- 74.70\%})   & 3187.51 (\textbf{-- 27.05\%})\\

    \hline
    \multirow{2}*{\tabincell{c}{App9\\train}} & before  & 2784.74 & 7771.63 & 12687.07 \\ 
    & after2     & 2498.84 (\textbf{-- 10.27\%}) & 6480.68 (\textbf{-- 16.61\%})   & 10863.83 (\textbf{-- 14.37\%})\\

    \hline
    \multirow{2}*{\tabincell{c}{App9\\predict}} & before  & 2756.00 & 7880.20 & 12578.81 \\ 
    &  after2     & 2485.14 (\textbf{-- 9.83\%}) & 6638.54 (\textbf{-- 15.76\%})   & 11175.23 (\textbf{-- 11.16\%})\\

    \hline
    \multirow{2}*{\tabincell{l}{App10}} & before  & 2697.58 & 4201.91  & 7471.82 \\ 
    &  after2     & 2505.34 (\textbf{-- 7.13\%}) & 3683.43 (\textbf{-- 12.34\%})   & 6573.01 (\textbf{-- 12.03\%})\\

     \hline
    \multirow{2}*{\tabincell{l}{App11}} & before  & 2688.60 & 7127.09  & 10245.73 \\ 
    &  after2     & 2420.26 (\textbf{-- 9.98\%}) & 5201.01 (\textbf{-- 27.02\%})   & 7945.03 (\textbf{-- 22.46\%})\\

    \hline
    \multirow{2}*{\tabincell{l}{App13}} & before  & 2928.96 & 915.75  & 4181.88 \\ 
    &  after2     & 2632.38 (\textbf{-- 10.13\%}) & 798.41 (\textbf{-- 12.81\%})   & 3735.14 (\textbf{-- 10.68\%})\\

    \hline
    \multirow{2}*{\tabincell{l}{App14}} & before  & 2767.26 & 1213.27  & 4285.15\\ 
    &  after2     & 2369.78 (\textbf{-- 14.36\%}) & 966.29 (\textbf{-- 20.36\%})   & 3678.36 (\textbf{-- 14.16\%})\\

\hline
    \multirow{2}*{\tabincell{l}{App16}} &before & 1905.50 & 504.45  & 2748.74\\ 
    &  after2  & 1753.29 (\textbf{-- 7.99\%}) & 383.82 (\textbf{-- 23.91\%})   & 2457.23 (\textbf{-- 10.61\%})\\

    \hline
    \multirow{2}*{\tabincell{l}{App17}} &before & 2063.34 & 202.69  & 2629.53 \\ 
    &  after2  & 1840.20 (\textbf{-- 10.81\%}) & 173.23 (\textbf{-- 14.54\%})   & 2376.29 (\textbf{-- 9.63\%})\\

    \hline
    \multirow{2}*{\tabincell{l}{App18}} &before&  1959.28 & 591.09  & 3096.21\\ 
    &  after2 & 1846.48 (\textbf{-- 5.76\%}) & 504.00 (\textbf{-- 14.73\%})   & 2865.85 (\textbf{-- 7.44\%})\\

     \hline
    \multirow{2}*{\tabincell{l}{App19}} &before&   1826.22 & 646.50  & 2795.05 \\ 
    &  after2 &  1687.23 (\textbf{-- 7.61\%}) & 556.60 (\textbf{-- 13.91\%})   & 2580.41 (\textbf{-- 7.68\%})\\
   
   \hline
   \end{tabular}
   \end{threeparttable}
\end{table*}


\begin{table*}[t]
  \footnotesize
 \centering
 \caption{The details of the newly added FaaS applications executed on Google Cloud Functions.}
 \label{tab:newdataGoogle}
   \begin{tabular}{l|l|l|l|l}
   \hline 
    \rowcolor{gray!50}\textbf{App ID} & \textbf{Name}  & \textbf{Version} & \tabincell{c}{\textbf{Size (MB)}} & \tabincell{c}{\textbf{LoC (k)}} \cr

   \multirow{2}*{\tabincell{l}{App20}} & \multirow{2}*{\tabincell{l}{google-requests~\cite{Test20}}}&before&  6.09 &41.75   \\ 
    & & after2 & 2.94 &  25.56  \\
    
    \hline
   
   \multirow{2}*{\tabincell{l}{App21}} & \multirow{2}*{\tabincell{l}{google-pandas~\cite{Test21}}}&before&  122.80 &420.93  \\
    & & after2 & 82.22 & 196.24 \\
    
    \hline
   
   \multirow{2}*{\tabincell{l}{App22}} & \multirow{2}*{\tabincell{l}{LDA-classifier~\cite{Test22}}}&before&  337.53 &606.34  \\
    & & after2 & 316.35 & 427.27  \\
   
   \hline
   \end{tabular}
\end{table*}

\begin{table*}[t]
\begin{threeparttable}
  \footnotesize
 \centering
 \caption{The cold start result\tnote{*} of App20 to App22 (originally developed for Google Cloud Functions) on both AWS Lambda and Google Cloud Functions.}
 \label{tab:RQ6App2022}
   \begin{tabular}{l|l|l|l|l|l}
   \hline 
    \rowcolor{gray!50}\textbf{App ID}&\textbf{Executed platform}&\textbf{Version}&\textbf{Preparation phase latency}&\textbf{Loading phase latency}&\textbf{Total response latency} \cr

    \hline

    \multirow{4}*{\tabincell{l}{App20}} & \multirow{2}*{\tabincell{l}{AWS Lambda}}&before  & 900.06  & 708.70   & 2082.54  \\ 
    & & after2    & 848.48  (\textbf{-- 5.73 \%}) & 608.90 (\textbf{-- 14.08\%})   & 1890.49 (\textbf{-- 9.22\%}) \\ \cline{2-6}
    &\multirow{2}*{\tabincell{l}{Google Cloud Functions}}&before & 2146.06 & 345.18 & 3308.11\\ 
    & & after2  & 2025.58 (\textbf{-- 5.61\%}) & 305.85(\textbf{-- 11.39\%})   & 2972.61 (\textbf{-- 10.14\%})\\

    \hline

    \multirow{4}*{\tabincell{l}{App21}} & \multirow{2}*{\tabincell{l}{AWS Lambda}}&before  & 1149.46 & 3585.32  & 5137.01 \\ 
    & & after2  & 1007.52 (\textbf{-- 12.35\%}) & 2972.49 (\textbf{-- 17.09\%})   & 4378.16 (\textbf{-- 14.77\%})\\ \cline{2-6}
    & \multirow{2}*{\tabincell{l}{Google Cloud Functions}}&before & 2476.64 & 4194.23  & 7011.91 \\ 
    & & after2  & 2112.53 (\textbf{-- 14.70\%}) & 3306.28(\textbf{-- 21.17\%})   & 5731.38 (\textbf{-- 18.26\%})\\

    \hline

    \multirow{4}*{\tabincell{l}{App22}} & \multirow{2}*{\tabincell{l}{AWS Lambda}}&before  & 1064.06 & 1677.95  & 2834.28 \\ 
    & & after2     & 981.71 (\textbf{-- 7.74\%}) & 1459.05 (\textbf{-- 13.05\%})   & 2547.58 (\textbf{-- 10.12\%})\\ \cline{2-6}
   & \multirow{2}*{\tabincell{l}{Google Cloud Functions}}&before&  2243.30 & 2912.45  & 5972.66\\ 
    & & after2 & 2097.86 (\textbf{-- 6.48\%}) & 2562.56(\textbf{-- 12.01\%})   & 5408.63(\textbf{-- 9.44\%})\\
    
    \hline

   \hline
   \end{tabular}
  \begin{tablenotes}
     \item[*] Performance results are in milliseconds.
  \end{tablenotes}
   \end{threeparttable}
\end{table*}

\finding{Our approach can be generalized to performance optimization of FaaS applications written in JavaScript languages (another widely used language in serverless computing) or executed on Google Cloud Functions (another popular serverless platform). The application code loading latency is reduced by up to 77.15\%, thus decreasing up to 32.20\% of the total response latency. The results have evidenced that our approach can be potentially generalized to a variety of FaaS applications with heterogeneous implementation and underlying serverless platforms.}

%% file: section/discussion.tex

\section{Threats to Validity}\label{sec:threats}

\noindent\textbf{Internal validity.} In our measurement study, we explore the possible root cause of the cold start overhead of FaaS applications. Since FaaS applications may be affected by resource allocation or the network of the serverless platform, the obtained latencies may lead to possible percentage bias in Fig.~\ref{fig:cold_analysis}. To alleviate this threat, it has been a common practice in the serverless computing literature to calculate the average latency of multiple runs to represent the general level~\cite{WangATC2018, yu2020characterizing, lee2018evaluation}. In our study, we follow the previous studies~\cite{WangATC2018, yu2020characterizing, lee2018evaluation} to run the experiments multiple times to report the average results.
For our measurement study and the experimental evaluation, we measure FaaS applications 20 times and then use the average value as the final comparable result of the performance.



In addition, in our study, we identify indispensable functions of FaaS applications by constructing the function-level call graph. The inaccuracy or incompleteness of the call graph may lead to missing some indispensable functions to cause application failure. To mitigate the threat, \toolName adopts the strategy of identifying as many indispensable functions as possible. Moreover, we design a mechanism for FaaS applications to fetch and execute optional functions in an on-demand loading way. In future work, we plan to design a more accurate code identification for FaaS applications while guaranteeing the correctness and effectiveness of FaaS applications.

\noindent\textbf{External validity.} 
We evaluate \toolName with real-world Python-based FaaS applications executing on AWS Lambda. This may lead to the limited generalizability of \toolName to FaaS applications written in other languages or executed on other serverless platforms. To mitigate this threat, we present and answer the research question, i.e., RQ6. The result shows that our approach can be generalized to performance optimization of FaaS applications written in JavaScript or executed on Google Cloud Functions, indicating the generalizability of our approach. For compiled languages like C/C++, although optimizations are already done at compile time, used standard libraries are also imported into applications~\cite{heo2018effective, quach2018debloating}. Similar to FaaS applications, applications may bear the performance burden of carrying all the features in the code with no way to disable or remove optional features. Reducing unnecessary feature-related code in compile phase means code size reduction, which is helpful for performance improvement of the program written in compiled languages~\cite{brown2019carve, sharif2018trimmer, qian2019razor}. Therefore, loading only indispensable code can also be applicable to performance improvement of FaaS applications written in compiled languages.

%% file: section/relatedwork.tex
\section{Related Work}\label{sec:relatedwork}


Serverless computing has been used in a wide range of software applications~\cite{ao2018sprocket,shankar2020numpywren,pu2019shuffling}, and thus attracted increasing attention from the SE community~\cite{wen2021characterizing, awsIdlelifetime, WenServerless21, eismann2021state, sampe2020toward, wen2023rise}. Some measurement studies~\cite{wen2021characterizing,awsIdlelifetime} have been presented to characterize the performance of serverless applications or platforms to help developers select the most appropriate one. To facilitate developers to develop their serverless applications, Wen \textit{et al.}~\cite{WenServerless21} uncovered 36 specific challenges that developers encounter in developing serverless applications. Performance-related challenges are also included. A comprehensive study about serverless applications~\cite{eismann2021state} was presented by SE researchers to show specific usage characteristics, where performance is also a reason for serverless adoption. In addition to the attention from the SE community, other communities like the Systems research community, the Networks research community, and the Services Computing research community have made significant efforts related to serverless computing in performance optimization~\cite{fuerst2021faascache, OakesATC18, AkkusATC18}, resource management~\cite{zhang2021caerus, palma2020allocation}, communication optimization~\cite{shillaker2020faasm, jia2021nightcore}, etc. In our study, we present an application-level code analysis approach to optimize the code of FaaS applications. This approach can be adopted by software developers to improve the cold start performance of FaaS applications.

According to the related studies~\cite{vahidinia2020cold, LambdaColdStart, OakesATC18}, there are two types of approaches to cold start optimization in serverless computing. The first one is to reduce the number of cold starts through a ``keep-alive'' or pre-warm policy. For example, major serverless platforms like AWS and Azure use a fixed ``keep-alive'' policy to retain the resources in memory for several minutes after a function execution~\cite{LambdaColdStart, AzureColdStart}. Although such a policy is simple and practical, it does not consider the actual invocation frequency and function patterns. Therefore, there are still many cold starts for most requests. Moreover, developers can easily identify this policy, causing them to keep resources warm by making frequent dummy invocations in advance. This practice exacerbates the resource waste problem. Another approach is to present new systems by designing data caching mechanisms~\cite{OakesATC18}, managing container runtime~\cite{AkkusATC18}, or setting and restoring snapshots~\cite{silva2020prebaking, DuASPLOS2020} in the underlying platforms. Specifically, SOCK~\cite{OakesATC18} cached interpreters and commonly used libraries in containers to reduce the cold-start latency, and they provided the lightweight isolation mechanism for serverless functions. SAND~\cite{AkkusATC18} applied the application-level sandbox runtime sharing to reduce the number of containers and, thus, container preparation latency. Silva \textit{et al.}~\cite{silva2020prebaking} proposed the usage of checkpoint/restore. The designed system prototype can restore snapshots of previously started functions' runtime instead of re-executing new cold starts. Similarly, Catalyzer~\cite{DuASPLOS2020} set checkpoints of critical paths and restored application and sandbox runtime in Google's gVisor~\cite{gVisor}. However, these optimization studies have modified underlying platform designs or sandbox runtime mechanisms; thus, it is difficult to apply in presented infrastructures on different platforms due to extensive engineering efforts, maintenance, and security problems. Differently, our approach effectively optimizes the cold-start latency at the application level, and it allows developers to improve the performance of their FaaS applications without any additional overhead. 

\section{Discussion}\label{sec:discussion}

\textbf{Accurate detection and optional code backup.} In our paper, we use static program analysis to construct the call graph, because dynamic analysis generally produces huge runtime overhead, which is inflexible and impractical in real-world applications~\cite{henderson2014make,azad2019less,quach2017multi}. In static analysis, an accurate detection may depend on complex program analysis methods~\cite{programAnalysis}, e.g., inter-procedural analysis and context-sensitive analysis. These methods establish the global state representation to capture all information of the whole program, which infers the relationships between code in a relatively accurate way. However, most FaaS applications are written in dynamic languages such as Python and JavaScript~\cite{serverlesscommunitysurvey1}. Due to the dynamic features of these languages, building accurate call relationships is an open problem~\cite{obbink2018extensible,haas2020static}, i.e., it is not possible to obtain 100\% accurate detection. Therefore, we adopt the optional code backup in this paper.


\textbf{Additional changes or efforts in other language prototypes.} For FaaS applications written in different languages, \toolName may consider some additional customization and optimization. 
First, some languages like JavaScript may have no magic functions like Python. Based on this point, we use language-specific functions in \textit{Application Entry Recognition} component to cover this situation to provide flexible customization capabilities for different language features. Second, for the optimization in function-level rewriting, this part needs to implement the on-demand fetching and execution mechanism for optional code. \toolName may require consideration of the variable scope of code execution under this mechanism for different languages, in order to guarantee the correctness and availability of FaaS applications.

\textbf{Overhead analysis of on-demand loading code.} We design an on-demand loading mechanism for FaaS applications to fetch and execute the separated optional functions. When executing the optimized FaaS applications, some optional functions may be required during the function execution. In this situation, on-demand loading overhead will increase the function execution time, decreasing the performance optimization space. However, this runtime cost does not affect the warm-start latency of serverless functions because it is a one-time cost if the container or VM of the serverless function is not released. More details can be found in Section~\ref{sec:rq4}. Moreover, this overhead is mainly in reading the lightweight file, about 100 ms, which is much smaller than the cold-start latency. In our study, \toolName has made the application loading latency reduce by 28.78\% on average. Thus, it is worth sacrificing a little execution performance overhead for greater overall performance optimization.

\textbf{Discussion about serverless platforms.} Indeed, in ideal conditions where the platform can cache everything effectively, the application code loading latency will not be an issue, and the proposed technique will not be needed. However, in practice, serverless platforms cannot cache everything. Specifically, each function instance or VM has a memory restriction. Moreover, function instances that execute serverless functions are constantly changing. Once new serverless functions are continuously deployed to the serverless platform, cached data in all function instances or VMs needs to be updated, which will cause high maintenance overhead.


\textbf{Serverless optimization plugin.} Serverless Framework's official website has presented the ``Serverless Optimize Plugin''~\cite{slsoptimizeplugin}, which can reduce the file size of AWS Lambda functions written in JavaScript. The main principle of this plugin is to separately bundle required dependency libraries for every serverless function~\cite{slsoptimizeplugin1}, i.e., excluding dependency libraries required by other serverless functions in the same application. Compared with our function-level, fine-grained code analysis approach, this plugin's analysis granularity is at the dependency library level, which is coarser-grained. Moreover, it is challenging to identify optional code relative to the FaaS application, missing opportunities for potential performance optimization.

\textbf{Potential use of \toolName.} In this paper, we propose \toolName to optimize the cold-start latency of FaaS applications. Besides, \toolName can also be potentially useful for other software development tasks, such as reducing deployment time. Existing work reported that developers often face the long deployment time of serverless applications and proposed an annotation configuration-based framework named Nimbus~\cite{chatley2020nimbus} to reduce the deployment package size of Java applications on AWS Lambda, thereby reducing the deployment time. Similarly, \toolName can load only indispensable functions and separate optional functions to reduce the code size and the deployment time.

%% file: section/conclusion.tex
\section{Conclusion}\label{sec:conclusion}

We proposed \toolName, an application-level code analysis approach, to load only indispensable code to optimize the cold start performance of FaaS applications. Specifically, \toolName identified the code related to application functionalities by constructing the function-level call graph, and separated other code (called optional code) from the application. The separated optional code can be loaded in an on-demand way to avoid the inaccurate identification of indispensable code causing application failure.
\toolName was implemented as the Python and JavaScript prototypes and evaluated with real-world FaaS applications. 
Results demonstrated that \toolName efficiently reduced the application code loading latency (up to 78.95\%, on average 28.78\%), thereby reducing the cold-start latency. As a result, the total response latency of serverless functions was decreased by up to 42.05\% (on average 19.21\%). 
Compared with the state-of-the-art, \toolName achieved a 21.25$\times$ improvement in reducing the average total response latency.